\documentclass[acmsmall]{acmart}
\AtBeginDocument{%
  }

\setcopyright{cc}
\setcctype{by}
\acmJournal{PACMHCI}
\acmYear{2025} \acmVolume{9} \acmNumber{7} \acmArticle{CSCW251} \acmMonth{11} \acmPrice{}\acmDOI{10.1145/3757432}

\usepackage{multirow}
\usepackage{xcolor}
\usepackage{subcaption}
\begin{document}

\title{Towards Designing Social Interventions For Online Climate Change Denialism Discussions}

\author{Ruican Zhong}
\email{rzhong98@uw.edu}
\orcid{0009-0004-7169-0675}
\affiliation{%
  \institution{University of Washington}
  \city{Seattle}
  \state{WA}
  \country{USA}
}

\author{Shruti Phadke}
\email{shruti.phadke@drexel.edu}
\orcid{0000-0002-0524-830X}
\affiliation{%
  \institution{Drexel University, Information Science}
  \city{Philadelphia}
  \state{PA}
  \country{USA}
}
\author{Beth Goldberg}
\email{bethgoldberg@google.com}
\orcid{0000-0002-8636-6702}
\affiliation{%
  \institution{Jigsaw (Google)}
  \city{New York}
  \state{NY}
  \country{USA}
}
\author{Tanushree Mitra}
\orcid{0000-0002-9507-6192}
\email{tmitra@uw.edu}
\affiliation{%
  \institution{University of Washington}
  \city{Seattle}
  \state{WA}
  \country{USA}}

\renewcommand{\shortauthors}{Zhong et al.}

\begin{abstract}
As conspiracy theories gain traction, it has become crucial to research effective intervention strategies that can foster evidence and science-based discussions in conspiracy theory communities online. This study presents a novel framework using \emph{insider language} to contest conspiracy theory ideology in climate change denialism on Reddit. Focusing on discussions in two Reddit communities, our research investigates reactions to pro-social and evidence-based intervention messages for two cohorts of users: climate change deniers and climate change supporters. Specifically, we combine manual and generative AI-based methods to craft intervention messages and deploy the interventions as replies on Reddit posts and comments through transparently labeled bot accounts. On the one hand, we find that evidence-based interventions with neutral language foster positive engagement, encouraging open discussions among believers of climate change denialism. On the other, climate change supporters respond positively, actively participating and presenting additional evidence. Our study contributes valuable insights into the process and challenges of automatically delivering interventions in conspiracy theory communities on social media, and helps inform future research on social media interventions.
\end{abstract}




\maketitle

\section{Introduction}
\begin{quote}
    \textbf{Climate Change Denial Post 1:} This is actually two horrible scandals rolled into one. Lyme disease has been suppressed and ignored for decades by the conventional medical complex in the US and Canada at the behest of the CDC most likely because chances are it was designed as a bioweapon at the Plum Island facility just off the coast of Old Lyme, CT [Connecticut] where it was seen for the first time, hence the name. \ldots Ticks are very hardy, and can live in many habitats, but in the US it's the eastern seaboard, mid-west, and west coast that are the most infiltrated, not the southern US where the climate is naturally warmer. So this claim that ticks can't live in colder climates and global warming is allowing them to expand their territory is entirely bogus, but suits two bogus agendas: ``climate change'' and the medical establishment's horrendous but deliberate handling of Lyme disease. I have to give them credit for trying to roll all these criminal actions into one though as it probably fools most people. (\textit{r/climateskeptics})
\end{quote}
\begin{quote}
    \textbf{Climate Change Denial Post 2:} They [The governments] are politicizing any fire to be about climate change now. They did it recently with the Amazon. They did it with the California fires, too. Basically, they are taking completely normal, \textbf{natural occurrences}, that happen every year for the past 1000 years, and rebranding them as proof of climate change. (\textit{r/conspiracy})
\end{quote}
\begin{quote}
    \textbf{Climate Change Denial Post 3:} Don't you feel a bit silly thinking a \textbf{trace gas} in the atmosphere could cause the climate to change and cause catastrophic events, none of which have ever happened. No \textbf{sea level rise}. only 1.3 degrees warming in 140 years (if \textbf{temperature measurements} since 1880 are even accurate) no increase in hurricanes, floods, droughts or wildfires, dramatic reduction in deaths caused by the environment, no climate refugees, no empirical evidence of cause and effect at all. (\textit{r/climateskeptics})
\end{quote}

These examples are only three of many conspiracy theories (CT) about how climate change is formed and how people should engage with climate change. Though large amounts of research have argued how the increasing amount of carbon dioxide is impacting global temperature\mbox{~\cite{lindzen1997can,abas2014carbon,graham1990increasing,montzka2011non}} and how the sea level has been increasing for the past decade\mbox{~\cite{mimura2013sea,cazenave2011sea,etkins1982rise}}, there has been continued disbelief of scientific evidence, especially in communities that are involved with climate change denialism.

With the rise in the attention for conspiracy theories (CT)\mbox{~\cite{douglas2015climate, kuzelewska2022rise}}, it has become important to explore mechanisms through which CT believers could be dissuaded from CT beliefs. Key characteristics of CT include 1) highlighting secretive actions of certain groups of people to achieve some end result\mbox{~\cite{dentith2018secrecy}} 2) claiming the existence of unethical behavior to achieve power, control, or other benefits\mbox{~\cite{moyer2019we}} 3) showing excessive distrust in official sources of information\mbox{~\cite{nera2022looking}} 4) demonstrating selectiveness of evidence or misinterpretation of evidence\mbox{~\cite{dentith2019conspiracy}} and 5) postulating that contradictory evidence is people's plot of the conspiracy itself\mbox{~\cite{uscinski2019conspiracy, uscinski2017climate, douglas2019understanding, douglas2017psychology}}. In the aforementioned examples from \textit{r/climateskeptics} and \textit{r/conspiracy}, the first example illustrates how climate change denialism believers consider that the claim of global warming is to cover up the occurrence of Lyme disease, demonstrating their belief in unethical behavior made by the medical establishments to achieve secret agenda. This corresponds to the first and second characteristics of CT. In the second example, climate change denialism believers showed a clear distrust in the government's claims and reports, which is related to the third characteristic of CT. The third example shows that climate change denialism believers cherry-picked evidence to make their claim and completely dismissed evidence about the rise in sea level from existing research\mbox{~\cite{mimura2013sea,cazenave2011sea,etkins1982rise}}, corresponding to the fourth characteristic of CT. In addition, their dismissiveness of the evidence of rise in temperature and natural disasters shows their tendency to ignore any claim that may contradict with what they believe in and consider that as part of the conspiracy itself, which is aligned with the fifth characteristic of CT. Overall, since the discussions in these climate change denialism communities demonstrate features of CT, we consider these communities as CT communities. To date, it has been challenging for researchers to interact with members of CT communities because of skepticism toward scientists and institutions \mbox{\cite{dunlap2011organized,painter2012cross}}. For researchers to study the effects of educational and pro-science messages or debunking evidence in online conspiracy discussions such as climate change denialism, it is important to explore communication strategies that allow researchers to maintain civil and mutually respectful conversations with CT community members.

How can we introduce interventions that motivate conspiracy theory believers to explore scientific inconsistencies and fallacies in conspiracy theories? Building on the previous work observing the role of doubt and deliberation in disengagement from online CT discussion~\cite{phadke2021characterizing,engel2023learning}, this paper contributes a theoretically motivated intervention design framework that can be deployed in open discussions on social media. Specifically, we investigate how insider language could be used to contest socially constructed conspiracy theory ideology. Insider language refers to terminologies that people use to discuss conspiracy theories that are unique to the community \mbox{\cite{phadke2021characterizing, phadke2022pathways}}. For instance, the bolded phrases in the examples above (e.g., \textit{``trace gas,'' ``sea level rise,'' ``temperature measurements,''} etc.) are some specific terms people often use in these communities to describe climate change-related concepts. In this work, we ask:

\begin{itemize}
    \item RQ1: How do CT believers (climate change deniers) respond to interventions contesting their beliefs?
    \item RQ2: How do climate change supporters respond in open discussions to interventions contesting climate change denialism?
\end{itemize}

Prior works in CSCW and the larger HCI community have long recognized the importance of bridging the gap between social requirements and technical capabilities\mbox{~\cite{ackerman2000intellectual,dwyer2007task,girardin2007towards,tuffley2009mind,treude2009tagging}}. However, the functioning of CT communities from this socio-technical standpoint remains understudied. Our work addresses this knowledge gap by examining both the social and technical infrastructures that shape participation in CT communities, specifically through the lens of members' responses to interventions. We highlight the social infrastructure by analyzing how different stakeholders—climate change deniers and supporters—utilize evidence and claims in their debates, and how interventions could influence these interactions. This reveals the underlying mechanisms of community dynamics and identifies potential avenues for effective intervention engagement. Our work also points out the technical infrastructure, including the tools, information sources, and policies that CT community members use in their interactions with other stakeholders and interventions. Building on existing research studying misinformation and CTs\mbox{~\cite{aghajari2023reviewing,baughan2022shame,engel2023learning}}, our study pinpoints the current disconnect between social and technical infrastructures and proposes strategies for designing interventions that effectively bridge this gap.

In our work, we first manually create intervention messages for the climate change denialism debate on Reddit by focusing on the insider language of the climate change denialist community. Focusing on in-group language or insider language used in the climate change debate allows us to contribute relevant intervention messages. We then use generative AI to expand our manually designed set of intervention messages. Next, we deploy interventions in the form of replies to posts or comments made by community members on Reddit using transparent Reddit bot accounts---accounts that openly self-identify as bots. Finally, we evaluate the climate change deniers' and climate change supporters' responses to our bots' interventions and present a thematic analysis capturing intervention responses by various types of users in the climate change debate on Reddit. 
 
We find that believers' (Redditors who believe in climate change denialism) responses to interventions were influenced by the presence of evidence, the emotional valence of the original post, and certain terminologies (sensitive terms associated with pro-climate change perspective) used in the interventions. For example, when a hyperlink to peer-reviewed evidence was included in the bot's reply (22 out of 49 interventions), climate change deniers engaged more positively, analyzing details and even sharing personal experiences. However, when interventions lacked evidence, climate change deniers often requested evidence or supported conspiracy theories instead. These findings suggest that evidence-based interventions received more positive responses from believers regarding climate change beliefs and conspiracy theories. Moreover, we encountered climate change supporters, who generally exhibited a positive attitude toward our interventions. Our interventions facilitated discussions among these individuals, enabling them to express pro-climate change perspectives and present additional evidence. Our work is one of the first to explore the automatic deployment of interventions in an open social media space--conspiracy theory communities. In addition, due to the ever-evolving and uncontrolled landscape of open online discussions on Reddit, we encountered several difficulties with this experimental study design, including adversarial responses from the community and account bans, which could inform future work designing social media interventions.


\section{Related Works}
In recent years, the proliferation of social media platforms has provided an unprecedented space for the dissemination and amplification of conspiracy theories. These platforms can enable the rapid spread of unverified information and the formation of online communities centered around CT beliefs \cite{cinelli2022conspiracy, goreis2020social}. Often, conspiracy theorizing can lead to real-world consequences such as inaction due to climate change denialism or harm to public immunity due to anti-vax attitudes \cite{douglas2015social}. In this paper, we focus on online discussions around climate change denialism and leverage the insider language in such online discussions to create interventions aimed at contesting the meaning-making process inside the climate change denialist communities. 

\subsection{Climate change denialism}
Climate change, a scientifically established phenomenon, faces persistent challenges from conspiracy theories that aim to discredit its reality or downplay its significance \cite{tam2023conspiracy}. While the majority (82\%) of the general public believe in global warming, there are still people who are skeptical (11\%) or dismissive (11\%) of the idea\mbox{~\cite{sixamerica}}. For example, deniers may claim that temperature data is inaccurate or unreliable, suggesting that there is no significant warming trend \cite{cook2020deconstructing}. Other climate change denial arguments reject the role of humans in climate change and instead point to natural forces, such as the influence of the sun and cosmic rays\mbox{\cite{cook2020deconstructing,cbs, hossain2023sun}}. The climate change debate in the United States in particular is influenced by special interest groups \cite{dunlap2011organized} and has affected discourse within the scientific community through contrarian narratives \cite{lewandowsky2015seepage}. More importantly, disbelief in climate change and the accompanying anti-science attitudes have posed significant challenges for public policy design and corrective action \cite{dunlap2011organized,painter2012cross}. In this work, we examine climate change denialism from the lens of conspiracy theories \cite{douglas2015climate} and analyze key insider language terms that contribute to the meaning-making process of climate change conspiracy theories. 

\subsection{Conspiracy theories and insider language}
Conspiracy theories often rely on the use of insider language or specialized terminology, contributing to the formation and reinforcement of group identity among believers \cite{hofstadter2012paranoid}. Conspiracy theorists may employ specialized language and coded symbols to communicate exclusivity among believers and establish boundaries between ``insiders'' and ``outsiders'' \cite{byford2011conspiracy}. Online platforms contribute to the creation and dissemination of insider language within communities~\cite{bessi2015science,lewandowsky2017beyond}. Researchers have analyzed how insider language in conspiracy theories may evolve across platforms \cite{phadke2021characterizing} and contribute to internet-mediated radicalization \cite{phadke2022pathways}. Given the important role played by insider language in conspiracy theorizing, in this paper, we leverage the insider language of the climate change denialist community to craft intervention messages that challenge the concepts behind the established insider language. 
This process is inspired by the theory of social interventions described below.

\subsection{Interventions for Conspiracy Theory Discussion}
Several research studies have experimented with interventions aimed at addressing conspiracy beliefs. These interventions often focus on strategies to mitigate the impact of conspiracy theories and reduce their influence on individuals. For example, research has explored the effectiveness of debunking conspiracy theories by providing factual information and evidence-based refutations \cite{chan2017debunking,benegal2018correcting}. However, studies have shown that mere exposure to corrective information may not always be sufficient \cite{lewandowsky2020debunking}. Other types of interventions also rely on inoculation theory. This approach involves pre-emptively exposing individuals to weakened forms of conspiracy theories or misinformation, followed by debunking information. Studies applying this theory have found success in strengthening resistance against conspiracy theories by providing individuals with tools to identify and refute misleading claims \cite{roozenbeek2022psychological}. However, inoculation applied to specific misleading narratives, such as climate change denialism, may not be as effective for audiences that already support those narratives, known as therapeutic inoculation~\cite{amazeen2022cutting}. The effectiveness of interventions can vary depending on individual differences, the nature of the conspiracy theory, and the underlying reasons for belief in conspiracies. Moreover, the majority of the intervention experiments are conducted in controlled environments as surveys or user studies. In this paper, we present one of the first works that deliver interventions in open, uncontrolled, and unstable social media spaces. Specifically, our interventions are inspired by the theory of social interventions which is described below. 

\subsection{Social Interventions}

The social intervention model from communication theory describes the dialogic social processes through which humans symbolically change the social systems \cite{brown1982attention}. The social intervention theory refers to attention, need, and power mechanisms that bring out changes in ideologies and social processes. We specifically focus on the ``attention'' mechanism that contests the socially constructed meanings to change ideologies. For example, in their original work on the attention model, Brown describes how changes, or attention shifts,  in the perception of Black identity in the early 1900s changed the racial relations in the United States \cite{brown1982attention}. 

Conspiracy theories are filled with contradictions that, when realized by the believers, can lead to a lack of belief and engagement in the conspiracy theory discussions \cite{phadke2021characterizing}. In this regard, drawing attention to the contradictions through dialogic social interventions may disrupt the conspiracy theorists' meaning-making process \cite{brown1982attention} and encourage conspiracy theory believers to critically reflect on their beliefs, potentially resulting in the goal of disengagement from the conspiracy community. In this paper, we craft interventions based on insider language of climate change denialism in open public discussions and analyze the responses by the community members.

\subsection{Countering Conspiracy Theories}
While a systematic approach to counter conspiracy theories has not been studied before, some prior works did touch upon a population that is involved in this process -- people who self-organize to counter misinformation and hate speech~\cite{prollochs2023mechanisms,drolsbach2023diffusion,birdwatch,dangerousspeech}. For instance, Drolsbach et al. discussed how community fact-checking can help address misinformation~\cite{drolsbach2023diffusion}. Similarly, Twitter/X launched Birdwatch where communities of fact-checkers could collaborate and crowdsource to flag and address misinformation~\cite{birdwatch}. Additionally, the Dangerous Speech Project exemplifies a resource where people volunteer to crowdsource resources against hate speech~\cite{dangerousspeech}. These existing efforts demonstrate that there are many people who actively engage and try to address misinformation and conspiracy theories in the communities. When we deployed our interventions in the communities on Reddit, we also encountered a group of users who were similarly putting in their efforts. While they did not work together, they actively engaged with posts in the communities individually, providing rationale and reasoning for the pro-climate change perspective. We discuss them in more detail in \autoref{section:results}. Our work corroborates the existence of this group of users, and highlights future design opportunities to engage with this group in \autoref{section:discussion}.






\section{Method}
\begin{figure*}[t]
    \centering
    \includegraphics[width=\linewidth]{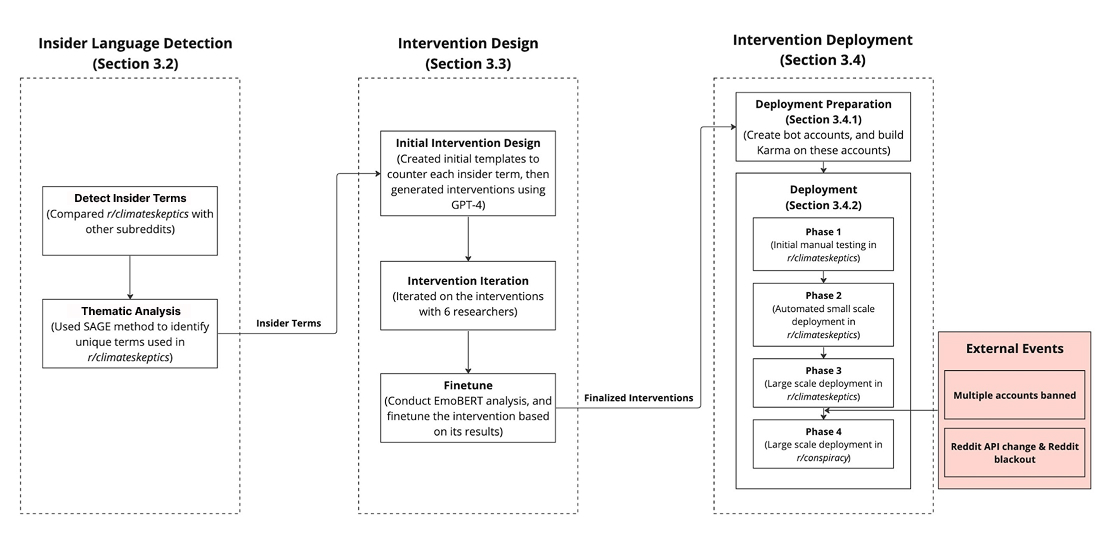}
     \caption{An overview of our intervention design and deployment process. The external events refer to events that occurred on Reddit that we could not control.}
    \label{fig:timeline}
\end{figure*}

In the following section, we discuss how we designed, implemented, and deployed our interventions to selected subreddits. We first used the SAGE (Sparse Additive Generative Models) method to analyze and identify insider language terms specific to the subreddits we are studying~\cite{eisenstein2011sparse, phadke2022pathways}. We then designed our interventions and iterated them through qualitative exercises to ensure their effectiveness. Finally, we deployed our interventions in the subreddits in three phases and collected community members' responses to our post. The entire steps are also shown in \autoref{fig:timeline}. 

\subsection{Subreddits containing climate denialism}
Our study focused on two subreddits: \textit{r/climateskeptics} and \textit{r/conspiracy}. \textit{r/climateskeptics} was created in 2008 and has about 43k subscribers. The community focuses its discussions on questioning climate-related environmentalism and sharing pieces of news and evidence to refute climate change science. Since \textit{r/climateskeptics} solely focuses their discussions on climate change conspiracy theories, this community was the main focus of our study. \textit{r/conspiracy} was created in 2008 and has about 2 million subscribers. The community discusses a wide range of conspiracy theory-related topics, including climate change and environmentalism. Prior research has studied these communities to identify conspiracy theory discussion dynamics and narratives~\cite{samory2018government, klein2019pathways,samory2018conspiracies}. These communities were also selected because they explicitly permit bots in their community rules and have a community norm accepting bots that post in discussions.

\begin{table}[]
\centering
\scriptsize
\caption{This table contains examples of the insider terms we used.}
{\renewcommand{\arraystretch}{1.35}
\begin{tabular}{|p{2.3cm}|p{10.8cm}|p{0cm}|}
\hline
\textbf{Insider Terms}                            & \textbf{Definition}                             \\ 
\hline
\textbf{aerosols}  & \begin{tabular}[c]{@{}l@{}} Opposite effect of greenhouse effect, so can be used to reverse global warming \end{tabular} \\ 
\hline
\textbf{albedo }  & \begin{tabular}[c]{@{}l@{}} albedo as a "natural" factor for climate change, and not GHGs\end{tabular}   \\ 

\hline
\textbf{black radiation}  & \begin{tabular}[c]{@{}l@{}}black radiation is a consequence of the atmospheric thermal effect rather than a cause for it, it allows for\\ both cooling and warming\end{tabular}              \\ 

\hline
\textbf{butterfly effect}  & \begin{tabular}[c]{@{}l@{}} butterfly effect is a consequence of the atmospheric thermal effect rather than a cause for it\end{tabular}                  \\ 

\hline
\textbf{ceres data}  & \begin{tabular}[c]{@{}l@{}} Many forms of questioning its authenticity, including through fringe research papers and funding bias\end{tabular}                  \\ 

\hline
\textbf{chaos theory}  & \begin{tabular}[c]{@{}l@{}} If everything is driven by chaos theory, how can one be sure that global warming is anthropogenic \end{tabular}\\ 

\hline
\textbf{climate alarmism}  & \begin{tabular}[c]{@{}l@{}} Some of the most extreme predictions made by climate alarmists have not yet come to fruition, leading\\ some to question the validity of their arguments, beware of ``alarmist talking points''\end{tabular}  \\ 

\hline
\textbf{consumer}  & \begin{tabular}[c]{@{}l@{}} Citing different forms of consumer exploration as motivation to advocate for climate change theory\end{tabular} \\ 
\hline

\textbf{freeze ray}  & \begin{tabular}[c]{@{}l@{}} Theory that warming gases can emit ``freeze rays'' and they can cool things down\end{tabular} \\ 
\hline

\textbf{gas bad}  & \begin{tabular}[c]{@{}l@{}} Rhetoric that climate change advocates just blandly say ``gas bad'' when GHGs are essential for sustaining\\ life on planet\end{tabular} \\ 
\hline

\textbf{glaciers melting}  & \begin{tabular}[c]{@{}l@{}} Glaciers have been melting for centuries\end{tabular} \\ 
\hline

\textbf{greenhouse effect}  & \begin{tabular}[c]{@{}l@{}} ``The reason we are able to even live on this planet at all is because of an `Atmospheric Radiative \\GreenHouse Effect' (rGHE) where so-called `GreenHouse Gases (GHGs)' – like CO2 – make it so that some \\of the energy that leaves the surface of the earth never manages to escape the system as a whole to space, \\but is rather `recycled' internally between atmosphere and surface, creating `extra' atmospherically induced\\ warming of the surface on top of the original solar warming.''\end{tabular} \\ 
\hline
\textbf{mid holocene}  & \begin{tabular}[c]{@{}l@{}} Repeated comparisons to how mid-holocene climate was warmer than current climate\end{tabular} \\ 
\hline

\textbf{radiative cooling}  & \begin{tabular}[c]{@{}l@{}} Radiative cooling is possible for surface temperature, any warming from greenhouse gases is offset by more \\radiative cooling from Earth's surface, ``New paper finds greenhouse gases causing radiative cooling, \\not warming, at current Earth surface temperatures''\end{tabular} \\ 
\hline

\end{tabular}}

\label{table:insider-terms}
\end{table}

\subsection{Insider Language Detection}
To ensure that our interventions were relevant and effective in \textit{r/climateskeptics}, we first investigated how the community discussed the climate change denialism perspective. We identified the ``insider terms'' of that community, which refers to the unique language used in the community. Prior research suggests that engaging people with these insider terms can be more effective in promoting critical thinking and challenging conspiracy theory logic~\cite{byford2011conspiracy,bessi2015science,lewandowsky2017beyond,phadke2021characterizing,phadke2022pathways}. To do so, we used the SAGE (Sparse Additive Generative Models) method, which has been adopted to extract and identify insider language in studying conspiracy theories\mbox{~\cite{eisenstein2011sparse, phadke2022pathways}}. SAGE is a method that uses a regularized log-odds ratio to contrast different text corpora's word distributions, highlighting the words that are uniquely used in each text corpora. In particular, we conducted a comparative analysis of r/climateskeptics against other subreddits (\textit{r/conspiracy}, r/\textit{conservative}) and pinpointed terminology distinctive to \textit{r/climateskeptics}. We excluded words that were generic, that could have other meanings and those that were not frequently used in the last three months. We then conducted open coding on the identified insider language terms (see: \autoref{table:insider-terms}) and highlighted how the community members share their insights and interact with one another. We then analyzed how these terms are used in anti-climate change context and brainstormed counter-arguments to these uses.


\subsection{Intervention Design} 
Using the ``insider terms'' identified and the counterarguments brainstormed, we crafted a bank of intervention messages. We then conducted brainstorming sessions with 6 researchers to revise the messages so that the interventions would be more polite~\cite{jary1998relevance,ryabova2015politeness}, and respectful~\cite{mackenzie2011communication}. We then conducted a thematic analysis of the revised interventions. As common themes emerged, we developed sample responses used as input to ChatGPT to expand our set of interventions by generating additional similar comments, leveraging the language models to adapt the intervention messages to suit the specific context and tone desired. We used ChatGPT to generate responses because we intended to test the scalability of LLMs in interventions. All generated responses were still validated by humans afterward to ensure their quality and relevancy. For instance, ChatGPT generated an intervention such as the following:

\begin{quote}\textit{There is a basic difference between mid-Holocene trends and current global trends. The mid-Holocene changes were much more gradual and over a 2000-year period, while current trends are at a much faster rate. This has been proven by researchers as well. https://www.rutgers.edu/news/important-climate-change-mystery-solved-scientists Do you think we should consider this faster rate of change to be a concern?}\end{quote}

In our human-validated edits, we added polite phrasings to make our interventions more approachable:

\begin{quote}
    \textit{\textbf{Thanks for sharing. You might consider} whether there is a difference between mid-Holocene trends and current global trends. The mid-Holocene changes were much more gradual and over a 2000-year period, while current trends are at a much faster rate. \textbf{There is some research on this you can check}: https://www.rutgers.edu/news/important-climate-change-mystery-solved-scientists \textbf{What is your take?} Do you think we should consider this faster rate of change to be a concern?}
\end{quote}

\begin{figure}[h!]
    \centering
    \includegraphics[width=0.85\linewidth]{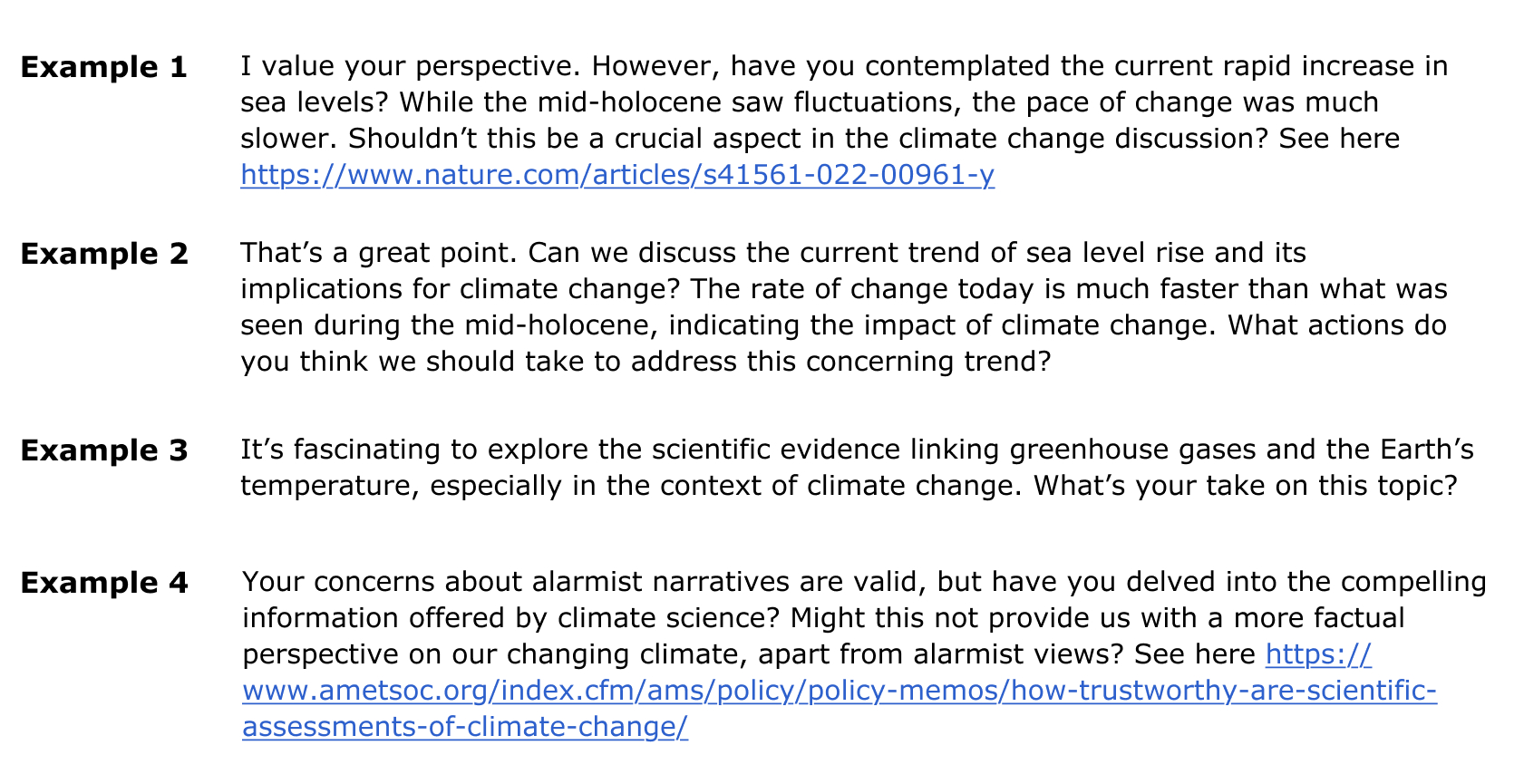}
    \caption{This figure contains four examples of the interventions deployed.}
    \label{fig:intervention-sample}
\end{figure}

Additionally, to ensure consistency and minimize chances for reactance, we wanted to make sure that our interventions were consistently expressing a positive attitude. To control for this, we used EmoRoBERTa, a pre-trained Speaker-Aware Emotion Recognition large language model, to identify the emotion of all the generated interventions. Specifically, EmoRoBERTa is a fine-tuned transformer model that could detect the emotion in text. Given text input, it is able to classify whether the writer was confused, engaged, curious, neutral, and so on. We used EmoRoBERTa to calculate the emotional scores of the interventions \mbox{\cite{kim2021emoberta}}. We then fine-tuned the interventions to make them demonstrate more positive emotions than negative ones. Examples of the interventions generated are shown in \autoref{fig:intervention-sample}.

\subsection{Intervention Deployment}
During this phase, we created 5 Reddit bot accounts to deploy the intervention, conducted pilot testing, and then deployed the interventions on two Subreddits. Below we outline the details of our deployment. 

\subsubsection{Bot Account \& Karma Building}
First, we created 5 Reddit accounts that we used to deploy the interventions. To automate the intervention delivery, we opted to use transparently labeled bot accounts, which allowed us to deliver multiple interventions automatically. To be considered as credible Redditors, we needed our accounts to meet age thresholds and had a sufficient amount of karma, which is ``a reflection of how much one's contributions mean to the community.''~\cite{karma} To curate sufficient karma so that the accounts could receive enough attention in the Reddit communities, we conducted Karma farming~\cite{rocha2023passive}. Karma farming is a common practice people use on Reddit to obtain karma points, where karma points reflects one's authority and establishedness in the community. We adhered to Reddit's policy and made posts and comments regularly in communities about gaming, such as \textit{r/FUTMobile} and \textit{r/HadesTheGame.} Specifically, we provided constructive comments about the games and conversed frequently with community members in these communities. Over a few weeks, we curated about 100 karma points for each account, which is a reasonable level for us to demonstrate our credibility. 

\subsubsection{Deployment}
We used the following approach to post our intervention messages to each subreddit community: when an insider term was mentioned in a post, a bot account would reply with a corresponding intervention message regarding that term. 
We deployed the interventions in three phases (\autoref{fig:timeline}). In the first phase, we tested if our deployment script worked accurately and if our deployment received responses from the community. Upon detecting comments that contained the insider language, our system would send email alerts about the posts. We would then review each instance and craft tailored responses to ensure that the intervention addressed the believers' points. This pilot test lasted three days to ensure that our intervention deployed effectively. In the second phase, we deployed the interventions at a larger scale using two bots that automatically selected prompts and deployed them in the community. To avoid posting repetitive or closely-timed comments from a single profile, we implemented a rotation system where the two automated accounts took turns performing the detection task at a 30-minute interval. In the final phase, we deployed our intervention at scale in \textit{r/climateskeptics}. However, we encountered multiple setbacks, including the Reddit blackout. During this process, we became too visible in \textit{r/climateskeptics} so we chose to redeploy on another conspiracy community. Thus, we proceeded with our deployment in \textit{r/conspiracy}. While our interventions were suitable for \textit{r/climateskeptics}, they were not as effective in this community because \textit{r/conspiracy} had slightly different insider language about climate change. We also encountered a few challenges with the Reddit platform throughout the deployment phase. We explain them in more detail in \autoref{appendix:lesson-learnt} as a reference for future work. Despite the challenges, we completed our deployment on Oct 6th 2023. After completing the deployment, we proceeded with data analyses to understand the challenges and opportunities in intervening in conspiracy theory communities.

\subsection{Data Collection \& Response Analyses}
\subsubsection{Data}
\label{section:data}

\begin{figure*}
    \centering
    \begin{subfigure}[t]{0.47\textwidth}
        \includegraphics[height=5.5cm]{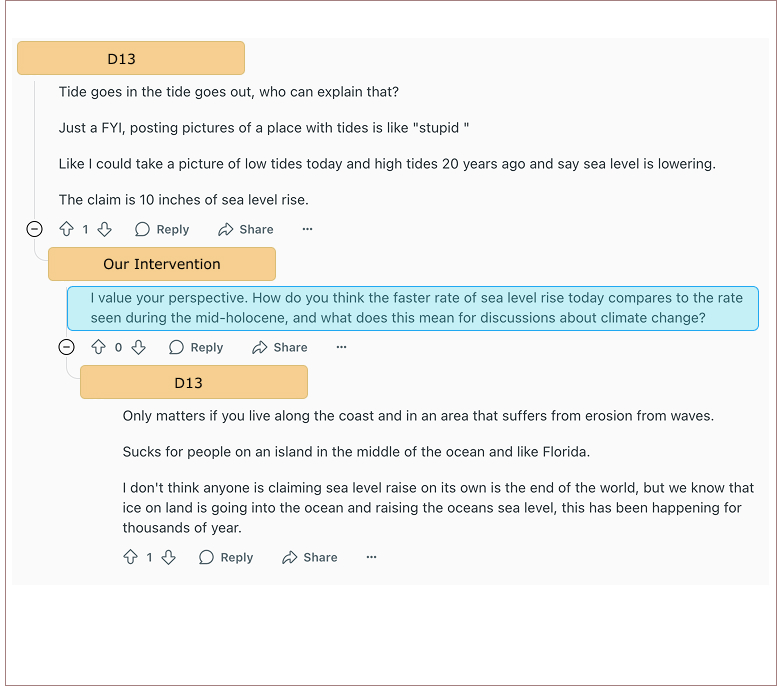}
        \caption{An example of an intervention deployed in \textit{r/conspiracy} that directly addressed the original poster's argument.}  
        \label{fig:interaction-example}
    \end{subfigure}
    \hfill
    \begin{subfigure}[t]{0.5\textwidth}  
        \includegraphics[height=5.5cm]{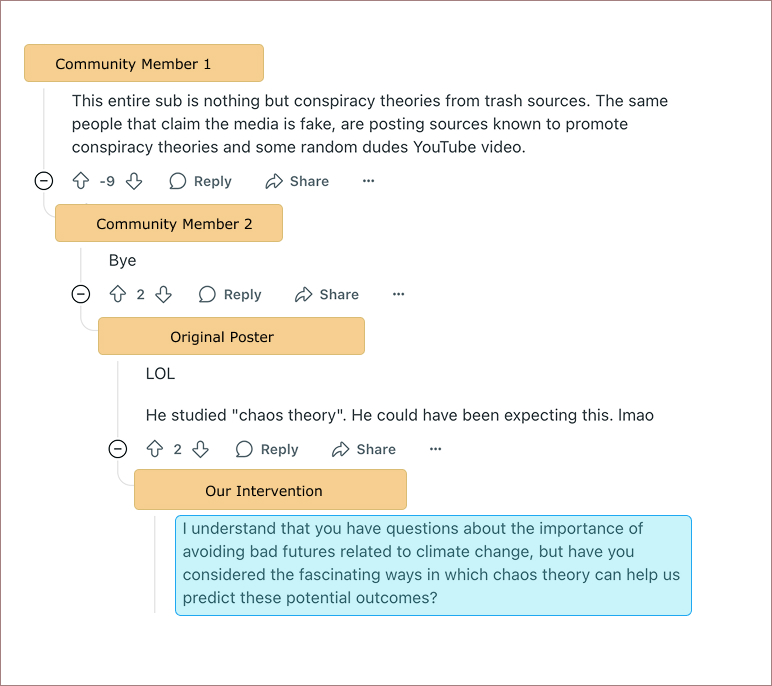}
        \caption{An example of an intervention in \textit{r/consipracy} that was out of context because the keywords detected were not associated with the climate change context.}    
        \label{fig:outofcontext-example}
    \end{subfigure}

    \vskip\baselineskip

    \begin{subfigure}[t]{0.47\textwidth}   
        \includegraphics[height=5.5cm]{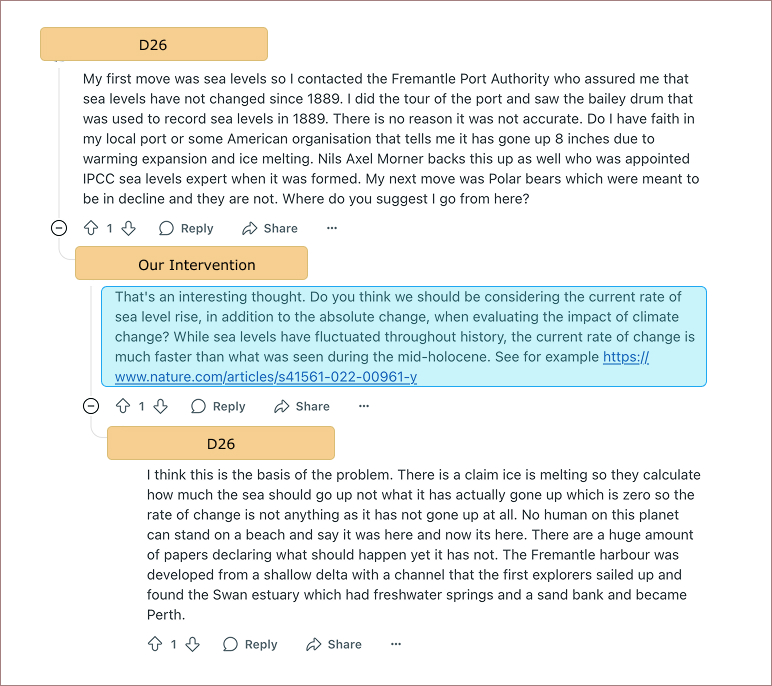}
        \caption{An example interaction between our intervention and the climate change deniers.}   
        \label{fig:believer-example}
    \end{subfigure}
    \hfill
    \begin{subfigure}[t]{0.5\textwidth}   
        \includegraphics[height=5.5cm]{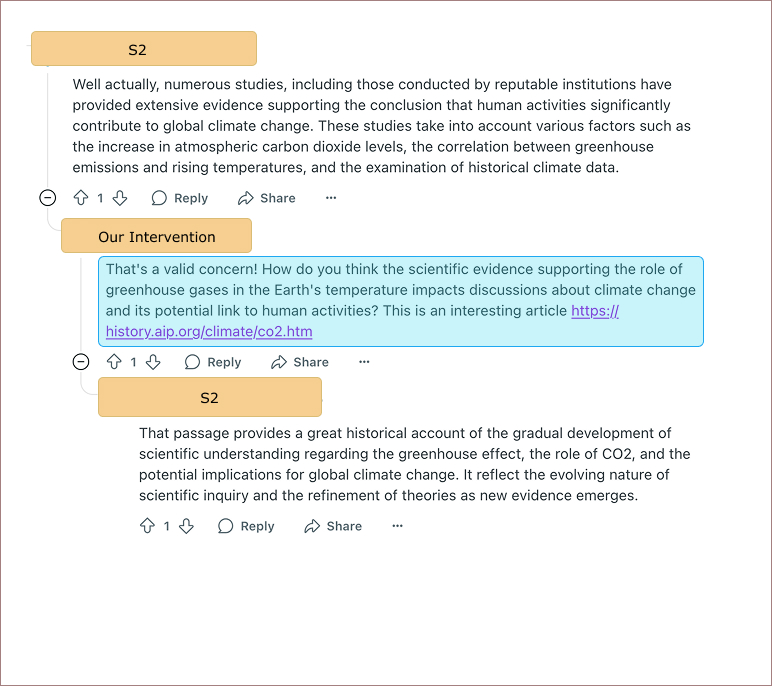}
        \caption{An example interaction between our intervention and the climate change supporters.}  
        \label{fig:bystander-example}
    \end{subfigure}

    \caption{This figure shows examples of how our intervention was deployed and how community members (including both climate change deniers and supporters interacted with our intervention.}
\end{figure*}

After the final deployment, we collected every response, including a total of 44 follow-up discussions in the thread. In total, we deployed 62 interventions on Reddit (46 were in \textit{r/climateskeptics}, 16 were in \textit{r/conspiracy}). 42 of these interventions received responses. 29 of the responses were by the original poster. \autoref{fig:interaction-example} shows an example intervention and people's response to it. Additionally, we observed that 36 of 46 interventions deployed in \textit{r/climateskeptics} were relevant. Some of our interventions misfired because the participants had been discussing different topics than the post that invoked the insider term. \autoref{fig:outofcontext-example} shows an example of an out-of-context example. While the original poster mentioned the term ``chaos theory,'' the discussions in the post was not about this in the context of climate change. Among the 16 interventions deployed in \textit{r/conspiracy}, 6 were relevant. As we mentioned in the previous section, this phenomenon was more prominent with our interventions deployed in \textit{r/conspiracy} because we built our interventions using insider language from \textit{r/climateskeptics}. These terms were slightly different from those of \textit{r/conspiracy}. We removed the out-of-context interventions from data analysis. In total, we had 42 data points for analyses.




\subsubsection{Analyses}
First, we analyzed all the original posts that we intervened on and identified two groups of Redditors that we interacted with. The first group are climate change deniers, who rejected scientific consensus on climate change. The other group are climate change supporters, who supported our post. (\autoref{fig:believer-example} \& \autoref{fig:bystander-example}). We describe the characteristics of each population in more detail in \autoref{finding:population}. Considering the original posters and all the Redditors who engaged in follow-up discussions, we interacted with 72 Redditors. Among these Redditors, 57 were climate change deniers, and 15 were climate change supporters. Among the 42 relevant interventions deployed, 30 were deployed to climate change deniers and 12 were deployed to pro-climate change users.

We conducted a thematic analysis of posts by each group separately. The authors developed a codebook by analyzing extracted excerpts from the responses to our interventions. Throughout the analysis process, one researcher first performed individual analysis and extracted excerpts. Then three researchers discussed the excerpts extracted until they decided on a final codebook. The final codebook includes themes such as \textit{climate change deniers' response to evidence-based intervention, climate change deniers' response strategies}, and \textit{climate change supporters' response strategies.} One researcher then used the codebook to code all the data collected. Throughout the coding process, the researchers analyzed climate change deniers' and supporters' posts separately. In the analysis of deniers' posts, we focused on understanding how the deniers responded to the interventions. Additionally, to understand any variation in emotional reactions that original posters and responders had to our intervention, we analyzed the collected data using EmoBERT\mbox{~\cite{aduragba2021detecting}}. Recall that EmoRoBERTa is a fine-tuned transformer model that could detect the emotion in text, classifying the writers' emotion when writing the text. Given text input, EmoRoBERTa provides granular emotion labels such as confused, engaged, curious, neutral and so on. To make it clearer for the analysis, we classified these labels into three categories: positive, neutral, and negative. For instance, engaged is considered as positive, curious is considered as neutral, whereas anger is considered as negative. Specifically, we contrasted the emotion between the original post and the response to our intervention to understand if our intervention had any impact on users' emotions in the discussion. We included the results of believers' posts in \autoref{fig:emobert-result}. In the analysis of the supporters' posts, we focused on how they interacted with our interventions, and how our interventions helped them reinforce their opinions in the community. We conducted the same EmoBERT analysis with pro-climate change Redditors' posts. However, we did not observe significant correlation as we did with believers' posts.

\begin{table}
\centering
\small
\caption{This table shows example responses to our interventions corresponding to each theme in our results. For instance, the second
line contains example responses to interventions where believers were emotionally invested and those where believers were more curious in their original post.}
{\renewcommand{\arraystretch}{1.17}
\begin{tabular}{|p{1.7cm}|p{2cm}|@{\hspace{4pt}}p{9.4cm}@{\hspace{4pt}}|}
\hline
\textbf{Stakeholder} & \textbf{Theme} & \textbf{Definition and Examples} \\
\hline

\multirow{3}{*}{\raggedright\small\textbf{Deniers}} & \raggedright\footnotesize\textbf{Evidence} & 
\colorbox{yellow}{\scriptsize\textbf{Intervention Contained Evidence:}}
\begin{quote}\scriptsize
\textit{\textbf{Intervention:}} ``I value your perspective. However, have you thought about the current rapid escalation in sea levels? While the mid-holocene saw fluctuations, the pace of change was much slower. Shouldn't this be a crucial factor in the climate change debate? See here [Linked Resource]''
\end{quote}
\begin{quote}\scriptsize
\textit{\textbf{Response:}} ``I'm sorry my friend I do not believe that sea levels are different \ldots stuff still looks just the same as it did when I was a kid.'' (D21)
\end{quote}

\colorbox{yellow}{\scriptsize\textbf{Intervention Contained No Evidence:}}
\begin{quote}\scriptsize
\textit{\textbf{Intervention:}} ``Do you believe that alarmist language around climate change could be a result of the growing body of evidence? Is it possible that these new terms help us better understand the issue?''
\end{quote}
\begin{quote}\scriptsize
\textit{\textbf{Response:}} ``There is no body of evidence presented here so the `new terms' only serve to obfuscate the fact.'' (D24) 
\end{quote}
\\
\cline{2-3}

& \raggedright\footnotesize\textbf{Original Poster's Emotion} &
\colorbox{yellow}{\scriptsize\textbf{Original Poster was Emotional:}}
\begin{quote}\scriptsize
\textit{\textbf{Intervention:}} ``Your concerns about alarmist narratives are valid, but have you delved into the compelling information offered by climate science? Might this not provide us with a more factual perspective on our changing climate, apart from alarmist views? See here [Linked Resource]''
\end{quote}
\begin{quote}\scriptsize
\textit{\textbf{Response:}} ``Evidence of climate change? \ldots Compelling bullshit is what this is.'' (D18) 
\end{quote}

\colorbox{yellow}{\scriptsize\textbf{Original Poster was Calm:}}
\begin{quote}\scriptsize
\textit{\textbf{Intervention:}} ``That's an interesting perspective. Do you think we should be considering the rate of sea level rise, in addition to the absolute change, when evaluating the impact of climate change? While sea levels during the mid-holocene were different, the rate of change was much more gradual than what we're experiencing today. See here [Linked Resource]'' 
\end{quote}
\begin{quote}\scriptsize
\textit{\textbf{Response:}} ``The arctic ice melt is further evidence of this. The rate of change is still very slow today. Certainly it's slow enough to adapt to it calmly.'' (D28)
\end{quote}
\\
\cline{2-3}

& \raggedright\footnotesize\textbf{Responses to Intervention with No Mention of Terminologies} &
\colorbox{yellow}{\scriptsize\textbf{``Alarmist'' was Mentioned:}}
\begin{quote}\scriptsize
\textit{\textbf{Intervention:}} ``Your caution against alarmist dialogue is noted, but have you explored the breadth of data that climate science offers? Could this not expand our understanding of climate change, leaving alarmist exaggerations aside?'' 
\end{quote}
\begin{quote}\scriptsize
\textit{\textbf{Response:}} ``Alarmist dialogue is wrong. Have you heard of a thing called the fossil record? In which extremes in climate are forever recorded? Let's include that in any discussion on climate change.'' (D27)
\end{quote}
\\
\hline

\multirow{2}{*}{\raggedright\small\textbf{Supporters}} & \raggedright\footnotesize\textbf{Responding with More Rationale} &
\begin{quote}\scriptsize
\textit{\textbf{Intervention:}} ``It's fascinating to explore the scientific evidence linking greenhouse gases and the Earth's temperature, especially in the context of climate change. What's your take on the topic?'' 
\end{quote}
\begin{quote}\scriptsize
\textit{\textbf{Response:}} ``Infarered energy escapes through water vapor. Water vapor isn’t the issue. It’s co2 and the methane and other harmful gases that us humans have unnaturally put into the atmosphere.'' (S12)
\end{quote}
\\
\cline{2-3}

& \raggedright\footnotesize\textbf{Responding with More Pro-climate Change Evidence} &
\begin{quote}\scriptsize
\textit{\textbf{Intervention:}} ``Your input is important! How does the scientific evidence supporting the role of greenhouse gases in the Earth's temperature impact our discussions about climate change and its potential causes? See this link [Linked resource]'' 
\end{quote}
\begin{quote}\scriptsize
\textit{\textbf{Response:}} ``Understanding greenhouse gases is essential to understanding the current climate crisis. [Linked resource]. Accelerated climate change is threatening the way we go about life \ldots Businesses will have to adapt quickly.'' (S12)
\end{quote}
\\
\hline

\end{tabular}}
\label{table:results}
\end{table}

\section{Results}
\label{section:results}




\subsection{Characterizing user population who engaged with our intervention}
\label{finding:population}
Our data analysis revealed that there were two groups of users who interacted with our intervention---\emph{climate change deniers} and \emph{climate change supporters}. In the following section, we discuss the characteristics of each group, including how they interacted with our post. 

\subsubsection{Interventions to Climate Change Deniers}
We define climate change deniers as those who oppose the opinions of our intervention and believe in anti-climate change. In particular, deniers expressed strong opinions about climate change scientists and the evidence they provided. For instance, D26 stated that \textit{``there is a claim ice is melting''} but that \textit{``no human on this planet can stand on a beach and say it was here and now it's here''} (\autoref{fig:believer-example}) to argue against sea level change. This is also a clear marker of climate change deniers. 

Upon analysis of the 40 relevant interventions deployed to the deniers, we found that 35 received responses. The average response length is 51.74 words (Min = 3, Max = 177). The average TF-IDF cosine similarity~\cite{christian2016single} between the response and our intervention is 0.54, demonstrating the relevance of people's responses. 

\subsubsection{Interventions to Climate Change Supporters}
We define climate change supporters as those who agree with the climate change perspective. Specifically, this group of people reinforces evidence supporting climate change. For instance, in \autoref{fig:bystander-example}, the original poster clearly states that a set of evidence contributes to global climate change, confirming the supporters' belief. Generally, the supporters agree with climate change stances, including the rising sea level and greenhouse gas emissions. They also cited evidence to support that stance. For instance, S12 cited external evidence to showcase that \textit{``the greenhouse gas emissions are heating up the planet.'' }
Among the 12 interventions deployed to climate change supporters, 3 did not receive a response and 9 received responses from the original posters. The average length of the responses to our interventions was 30 words (Min = 4, Max = 142). The TF-IDF cosine similarity~\cite{christian2016single} between the response and our intervention is 0.52. 

\subsubsection{Difference between Climate Change Deniers' and Climate Change Supporters' Responses to the Interventions}
In addition, to understand the differences between how climate change deniers and supporters responded to the interventions, we performed a t-test on the sentiment of the responses by the two groups. Our results indicated that there is no significant difference ($t(38) = 1.03, p > 0.05$). However, this could possibly be due to the small sample size we had, including 30 climate change deniers' responses and 10 supporters' responses.

\subsection{RQ1: How do climate change deniers respond to our interventions?}
We found that climate change deniers showed a more positive attitude in their responses when our intervention contained evidence, when the original poster was less emotionally invested, and when our intervention did not contain certain terminologies on climate change. We share examples of each theme in \autoref{table:results}.

\subsubsection{Responding depending on whether the intervention contains evidence}
19 of the 49 interventions deployed to climate change deiners contained evidence. 13 of the 19 original posters showed a positive attitude and responded to the details in the source. For instance, after D2 reviewed the evidence we shared, they described that the article was making a false claim about the sea level change because the article’s reasoning was \textit{``it can’t be explained by some other reason like subsidence.''} From D2’s perspective, the article \textit{``used the weasel words `could' and `most likely' which means like most other issues with climate change it is all speculation.''} Similarly, in an exchange with D24, we found that the deniers read deep into the linked article, and picked out a single detail to demonstrate how \textit{``the alarmists are hiding in the details and thinking we might not notice.''} While our intervention was unable to change climate change deniers’ opinions on the issue, introducing the article did motivate them to engage with our post. And in some situations, the deniers started sharing their personal experiences. When our intervention introduced evidence about the decades of empirical data by climate scientists, D42 started engaging by speaking about their prior experiences with this set of evidence, \textit{``in a previous life I worked on light and heat transfer measurements. Unfortunately, for the very simple experiment we were performing, the models did not come close to what we measured.''} In addition to the example, D23 further highlighted the boundaries of the framing of climate change debate in his/her terms, and how it would be different from that of our intervention: \textit{``I’m sure you can get a great (academic) career out of making climate change model, unfortunately, it has very little to do with science if you don’t have a way to validate your results.''} In this exchange, D42 shared more personal thoughts about why not believe in climate change as an establishment of the argument, especially regarding how current evidence may not be convincing enough in the deiners’ perspectives. 

In contrast, when no evidence is introduced in our intervention, climate change deniers directly asked for more evidence or choose to respond with conspiracy theory supporting evidence. In D2's response to our intervention without evidence, D2 explicitly said \textit{``based on what evidence''} and shared that \textit{``there are 400,000 glaciers in the world and only a handful have been studied completely.''} D2's response fortified climate change deniers' own theory and refuted our pro-climate change intervention as \textit{``speculation without evidence.''} (D2) These examples demonstrate that including evidence in interventions would help avoid deniers' fixation on the existence of pro-climate change evidence but instead help them focus their attention on having an open debate about climate change.



\subsubsection{Responding depending on whether intervention contained certain terminologies}

We found that when our intervention mentioned certain terminologies of climate change (e.g., ``alarmist,'' ``climate scientists''), the deniers (8/57) often responded in an emotional manner. For instance, when the term ``climate scientists'' was mentioned in our post, D11 responded: \textit{``So `climate scientists' lie to us, on purpose, because humans are too dumb to understand the mythology they're peddling? And that's OK with you? They do, by the way, deliberately lie and falsify data, it's very well documented. So forgive me for being skeptical about a group of proven liars and con-artists.''} In contrast, if our intervention just responded to the topic discussed instead of promoting the conversation to involve terminologies, the deniers showed a more positive attitude. When our intervention simply expressed a concern building on the original poster's argument and said \textit{``How do you think the faster rate of sea level rise today compares to the rate seen during the mid-Holocene, and what does this mean for discussions about climate change,''} deniers did not immediately shut down to discussion. Instead, D26 responded to our question by redefining \textit{``the claim ice is melting''} as the basis of the problem. B26 then continued to introduce the rationale that \textit{``there are a huge amount of papers declaring what should happen yet it has not''} to showcase how the pro-climate change perspective may fail here. In extension, D26 brought up an example of how \textit{``the Fremantle harbor was developed from a shallow delta with a channel that the first explorers sailed up and found the Swan estuary which had freshwater springs and a sand bank and became Perth.''} In this exchange with our intervention, the deniers showed a more open attitude of sharing their own thoughts and their own set of evidence of why they believed in the conspiracy theories instead of fixating on the specific terminologies and elevating the discussion to an emotional level.

\subsubsection{Responding depending on original poster's emotion}
\label{section:result-emotion}
\label{section:term}
\begin{figure}[h!]
    \centering
    \includegraphics[width=0.65\linewidth]{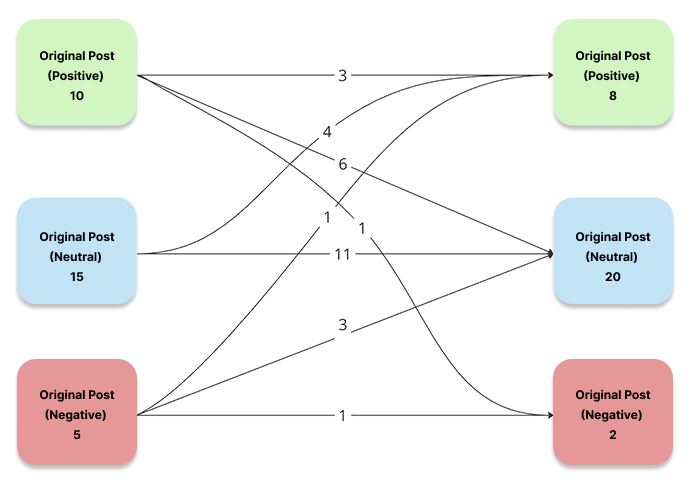}
    \caption{This image shows the EmoBERT analysis result of the comment (original post) and the response to the corresponding intervention. For the original post and its corresponding responses, we grouped the analysis result into three categories: positive (e.g., optimism, curiosity, etc.), neutral, and negative (e.g., anger, annoyance, etc.). We considered the negative posts as a reflection of the deniers showing a strong emotional investment in the conspiracy theories. In contrast, if the posts expressed a neutral or positive attitude, we considered that the deniers expressed a low emotional investment. Note that this graph only showcases the analysis of climate change deniers' original posts.}
    \label{fig:emobert-result}
\end{figure}
Using EmoBERT to test our data's emotional level (\autoref{fig:emobert-result}), we found that if deniers (15/57) expressed a neutral and/or curious attitude towards climate change denialism instead of deeply invested in them, they would be more likely to engage in a back-and-forth knowledge exchange with our bot account. And after the interaction with our intervention, 11 out of the 15 deniers (who showed a neutral and/or curious attitude) maintained their neutral and/or curious attitude, demonstrating that our interventions kept them open to conversations and further engagement. Additionally, 4 out of the 15 deniers expressed a more positive attitude after the interaction, showing a possible increase in how open they are to evidence and further conversations. For instance, when our intervention showed the \textit{``current rate of sea level rise''} with linked evidence, D26 who was asking questions about how to think about the \textit{``polar bears which were meant to be in decline and they are not,''} and asked \textit{``how do I think about this problem''} responded to our evidence with a well-defined discussion. Similarly, in an exchange with D2, we found that when the deniers held a more curious attitude, they are more likely to engage in an intellectual conversation. When D2 cited evidence explaining no increase in natural disasters causally lead to no climate change, D2 used the phrase \textit{``Can you show me where he is wrong?''} When we provided linked evidence, D2 responded by engaging with the citation, explaining how our evidence failed to address the issue: \textit{``Yearly changes in sea level are below the measurement accuracy of the instruments. This citation is saying ice is melting at the fastest rate and yet no one can point to any place on earth.''} Though D2 was still entrenched in climate change denialism, D2 was willing to engage in a discourse with our post. 

Based on \mbox{\autoref{fig:emobert-result}}, in most cases, deniers' emotional state improved or remained the same. However, one case came to our attention where the denier's emotion changed from positive to negative after encountering our intervention. Initially, D40 was sharing a detailed explanation about how the rise in population caused the temperature rise instead of climate change. In our intervention, we mentioned that \textit{``There is a consensus among scientists that greenhouse gases play a significant role in climate change.''} Then D40 responded with \textit{``There is no consensus, only bought-off scientists.''} This reinforced our result discussed in \mbox{\autoref{section:term}} that when intervention contained certain terminologies, deniers may be triggered and respond emotionally.

\subsection{RQ2: How do climate change supporters respond to our interventions?}
Our analyses did not show any direct correspondence between the original post, our intervention, and the corresponding responses (e.g., whether the intervention contained evidence, whether the original poster was emotional, or whether the intervention contained certain terminologies). But we did observe that climate change supporters generally showed a positive attitude toward our intervention. Our intervention allowed them to more easily discuss their pro-climate change perspectives and offered them the opportunity to provide more evidence. Examples of climate change supporters' responses are shown in \autoref{table:results}.

\subsubsection{Responding with more rationale}
Climate change supporters in the community used our intervention as a way to extend the discussions of the pro-climate change perspective. When S12 was arguing for the green movement, our intervention introduced the topic of greenhouse gases' role in the Earth’s temperature and climate. S12 then asked the following question: \textit{``Tell me what you know about greenhouse gases.''} The response demonstrated S12’s willingness to share more thoughts on the given topic. Furthermore, when we introduced pro-climate change evidence, S2 built on our intervention and discussed how the article explained the greenhouse effect. Using phrases such as \textit{``that passage provides a great historical account of,''} S2 shared additional insights about how new evidence such as the article \textit{``reflects the evolving nature of scientific inquiry and the refinement of theories.''} Similarly, when we introduced evidence about greenhouse gases and a climate change denier (D52) in the community responded with \textit{``There's no consensus. Only among bought-off scientists.''} S5, a pro-climate change Redditor, stepped in and argued in our favor: \textit{``No there's a consensus. Whether the consensus is 97\% of climate scientists or 88\% that's still a very strong consensus.''} In this interaction, our intervention received both climate change deniers' and supporters' attention, which encouraged more discussions and debates in the community. 

\subsubsection{Responding with more pro-climate change evidence} 
\label{section:term}
Our intervention also allowed climate change supporters to offer more evidence in follow-up discussions. When our response highlighted ``alarmist messages'' in response to a pro-climate change argument around temperature change, S8 responded with \textit{``The situation is alarming. We should all be alarmed''} followed by a list of detailed examples of temperature change, such as \textit{``Canada’s wildfire as of today have already burned more land than in any previous wildfire season,''} and \textit{``Texas was under a 45-degree Celsius heat dome for two weeks, killing dozens of people.''} In this interaction, the climate change supporters built on our intervention to reinforce their own argument. Similarly, when our intervention corroborated S10's post on sea levels are rising by asking the question of \textit{``Can we discuss the current trend of sea level rise and its implications for climate change? The rate of change today is much faster \ldots indicating the impact of climate change.''} S10 showed a positive attitude and shared more evidence supporting how \textit{``ice levels are connected.''} Further, as cited sources are provided, N10 also followed up with descriptions and rationales: \textit{``the ice caps reflect some sunlight in the visible range. As there is less ice, less gets reflected and so total warming increases.''} Overall, our engagement allowed the climate change supporters to provide more evidence and reinforce their argument in the climate change debate. 

\section{Discussion} 
\label{section:discussion}
In this work, we studied how to intervene in conspiracy theory communities and how these interventions may influence the way community members engage in climate change debate. Our work took a novel approach to design and deploy the interventions. By interacting with two groups of users, climate change deniers and climate change supporters, we identified what types of interventions worked well in these communities. We also gained insights about how climate change supporters engage in these conspiracy theory communities, pointing out how future work could study and engage with this population. 

\subsection{Designing Conspiracy Theory Interventions}
\subsubsection{Community-centric Interventions}
In online communities such as Reddit, the community as a whole shapes the way that people speak about the conspiracy theory and how they interact with one another around it. Many existing platforms such as Twitter/X has focused on engaging with the spread of misinformation from an individualistic perspective, where each account and each post is addressed as misinformation.
Furthermore, many existing works studying the design of social interventions focus on targeting individual users as well ~\cite{aghajari2023reviewing}. Our work instead took an approach to engage the community as a whole because the social context (where people encounter conspiracy theories and how others engage with them in that context) is an important aspect contributing to intervention design~\cite{hofstadter2012paranoid, byford2011conspiracy,bessi2015science,lewandowsky2017beyond, aghajari2023reviewing}. We demonstrated the possibility and importance of thinking about groups of conspiracy theorists as communities. We tried to design our interventions as closely related to the community as possible by engaging with insider language unique to that community. We also used GPT-4 to generate variations of the interventions so that the community members would not feel like they were engaging with an automated account. Additionally, we attempted to simulate the intervention experience as close to interacting with a real member of that community as possible because of the motivated reasoning phenomenon. Prior works suggest that the motivated reasoning phenomenon may occur if the conspiracy theorists experience an identity threat based on the provided evidence\mbox{~\cite{flynn2017nature,kahne2017educating,kunda1990case,kahan2013ideology,lodge2013rationalizing}}. In that case, conspiracy theory believers are more likely to engage their defense mechanism and choose to continue holding their original belief. Our approach engaging with insider language helped us better align with believers' identity and thus avoid the motivated reasoning effect. Our focus on insider language not only enabled us to meet the community where they were, but also helped us understand the discussion themes in these climate conspiracy communities. Overall, our results indicate the benefits of using community-centric interventions to engage with conspiracy theories. However, since our approach used a manual approach to collect and counter the insider terms, future work could explore how to automatically detect and analyze these insider languages unique to each community. This could make the process of designing interventions easier and more efficient.

\subsubsection{Evidence-based Interventions}

Providing evidence to support the pro-climate change perspectives was an important component of our intervention. Deniers showed a higher tendency to respond and engage when evidence was provided. However, existing literature pointed out that these communities may be prone to having scientific argumentation~\mbox{\cite{wofford2022parasitic,wofford2023curating}}, engaging in evidence provided by others through certain rationales to support their own claims. Future work could deploy interventions in a wide range of conspiracy theory communities to further understand the generalizability of the approach to use evidence. Additionally, when designing our interventions, we carefully considered how to best include the evidence since prior works suggested that the backfire effect~\cite{anderson1980perseverance,garrett2013promise,nyhan2010corrections,moravec2018fake,nickerson1998confirmation} could impact the conspiracy theory believers' consumption of information. As existing works noted~\cite{anderson1980perseverance,garrett2013promise,nyhan2010corrections,moravec2018fake,nickerson1998confirmation}, the backfire effect occurs when people fail to consume information objectively, which means they are less likely to change their opinions about the original information. Thus, we fine-tuned our interventions to be as respectful and polite as possible when raising opposing opinions and evidence. Future works intervening in these communities should also account for the possible impact of backfire effect when introducing evidence.

\subsection{Intervening climate change deniers}
In our study, we deployed our interventions in subreddits and observed how users interacted with our posts. After analyzing the collected data, we observed that there are two groups of Redditors in the community: climate change deniers and supporters. 

Our results indicated that when interacting with climate change deniers, the most important component of an intervention is linked evidence. When climate change deniers engaged with our interventions, they directly engaged with the provided evidence. In fact, deniers often explicitly ask for evidence if someone is proposing an opposing view. Without the evidence in the intervention, more back-and-forth discussions about what the evidence is and where the evidence is from would ensue. Our results echo prior research in other domains (e.g., health, psychology, education, etc.)~\cite{kratochwill2003evidence,kegeles2012intervention,albarqouni2018evidence}, showcasing that evidence is a key component in communicating ideas with others, especially if the goal is to intervene others' perspectives and the evidence is introduced in a respectful way. 

When deploying the interventions, another important aspect is to avoid certain terms such as ``climate change scientists.'' Explicitly mentioning the pro-climate change perspectives could lead to a complete shutdown towards any evidence and rationale posted in an intervention. Additionally, we found that how emotionally invested the original poster was, influenced how the poster responded. As explained in \autoref{section:result-emotion}, when the emotional investment was higher, there was a higher likelihood that the person would respond negatively. We also observed that climate change deniers were willing to share their personal stories to corroborate their point of view. Prior research has shown that personal stories can invoke emotional resonance with the audience~\cite{gargiulo2006power}. Leveraging existing discussions about how to share a personal story \mbox{\cite{lynn2018communicating}} and highlight human values \mbox{\cite{corner2014public}} in response to conspiracy theories, future research could explore how to best use an automated system to generate such interventions. Additional research could also investigate how to increase people's exposure to alternative points of view. For instance, an automatic system could be developed to respond to conspiracy theorists even after they started reacting emotionally. This approach could collect more insights about how conspiracy theorists engage in an emotional state, and help explore how to best engage with them in these situations.
Moreover, our work indicated that whether the conspiracy theory believers are emotionally invested is an indicator of how they would respond to interventions. Prior works have also suggested that ``emotional selection'' influences how people consume content\mbox{~\cite{sunstein2009conspiracy}}. Specifically, conspiracy theory believers are more likely to engage with content that correspond to their emotional state. Similarly, Heath et al. suggested that emotion snowballing could occur when intense emotion is spread from one another\mbox{~\cite{heath2001emotional}}. To more closely investigate how emotion influences people's perceptions of the interventions, future work could conduct experiments to compare how people respond across situations of emotional selection, emotion snowballing and so on. Another aspect worth exploring is how to sway the ``movable middle.''\mbox{~\cite{prooijen2018psychology}} The ``movable middle'' are people who have not decided about whether they would believe in the conspiracy theories yet. Future work could study this population specifically and design the interventions more tailored to their perspectives.

\subsection{Interacting with climate change supporters}
Prior studies have identified the existence of a group of users who actively counter misinformation and hate speech online~\cite{prollochs2023mechanisms,drolsbach2023diffusion,birdwatch,dangerousspeech, koo2021motivates}. Our work highlighted a similar group of users who actively engage with conspiracy theories. Specifically, our work revealed how pro-climate change Redditors interacted with members of conspiracy theory communities and how our interventions supported their efforts by providing additional evidence for climate change. Our work underscored that people who do not hold the belief of the community may play an important role in encouraging more open discussions in the community. Future work can more specifically focus on counter-attitudinal populations, or pro-climate science populations, and explore how to best support them in engaging effectively in conversations in conspiracy theory communities. They could explore the specific needs of these users, including how to support their process of collecting information, how to support their process of organizing that information, and how to support their process of interacting with conspiracy theory believers. For instance, based on Twitter/X's Birdwatch project and the Dangerous Speech Project~\cite{birdwatch,dangerousspeech}, there is an interesting opportunity to explore how to aggregate a central hub of information countering conspiracy theories to help alleviate these users' burdens.

\section{Ethical Considerations}
Throughout the design and deployment of the study, we tried to minimize harm. Specifically, we were interested in engaging with conspiracy theory community members to explore whether and how to change their beliefs. To do so, we needed to build trust with the community members. However, this is difficult to do because these communities do not trust scientists at all, which is one reason why they believe in conspiracy theories. Thus, instead of running a full deception study, we chose a middle ground. Building on existing research deploying interventions to misinformation and conspiracy theories \mbox{\cite{freiling2023science,munger2017tweetment,pennycook2020fighting,baughan2022shame,hangartner2021empathy,siegel2020no2sectarianism}}, when we created the accounts for deployment, we made sure to self-disclose the accounts as bots used for research purposes instead of deceiving the community members. However, since we did not explicitly disclose our accounts as bots when we posted interventions each time (the information that we were researchers was only listed under the account profile), some community members may not have realized that we were conducting research. We believe that future research should also ensure transparency using approaches such as account labeling when deploying interventions. When we collected data from the Reddit communities and used the SAGE method to identify insider language, we made sure all data was anonymized. When we analyzed the collected data (responses to our interventions), we ensured that all users were anonymized. We also made sure that we only collected and analyzed users' posts in the community and did not collect any de-identifiable information from them.  Second, when we designed our interventions, we also tried to fine-tune them to be polite and respectful. 

\section{Limitations}
There are some limitations to this work. First, our sample size was relatively small due to several external challenges we faced in the deployment phase, such as the Reddit blackout. Additionally, while lots of our interventions received upvotes and downvotes from community members, some did not receive a response in the text format. Thus, we had a relatively small sample to perform qualitative analysis on. Future large-scale deployment could help resolve these issues and provide more quantitative insights. Additionally, our work only explored how people responded to our intervention. We did not further engage with people who responded with follow-up conversations. Future work should study these follow-up interactions more closely to further understand how to best engage with conspiracy theory believers.



\section{Conclusion}
In this work, we designed and deployed interventions addressing conspiracy theories in Reddit communities. We designed interventions tailored to the conspiracy communities by addressing the specific insider terms they used. Our work highlights the process and challenges of automatically intervening in conspiracy theory communities. Additionally, throughout the deployment, we encountered a group of users who voluntarily engaged with conspiracy theories, highlighting future opportunities to study and engage with them. Overall, our work provides implications for future works studying and interacting with conspiracy theorists.

\section{Acknowledgement}
Research reported in this publication was supported by Google Jigsaw. We would also like to thank Neelesh Azad for helping us with the experimental design that supported the methodology write-up of the paper.





\bibliographystyle{ACM-Reference-Format} 

\bibliography{sample-base}


\begin{thebibliography}{91}


\ifx \showCODEN    \undefined \def \showCODEN     #1{\unskip}     \fi
\ifx \showISBNx    \undefined \def \showISBNx     #1{\unskip}     \fi
\ifx \showISBNxiii \undefined \def \showISBNxiii  #1{\unskip}     \fi
\ifx \showISSN     \undefined \def \showISSN      #1{\unskip}     \fi
\ifx \showLCCN     \undefined \def \showLCCN      #1{\unskip}     \fi
\ifx \shownote     \undefined \def \shownote      #1{#1}          \fi
\ifx \showarticletitle \undefined \def \showarticletitle #1{#1}   \fi
\ifx \showURL      \undefined \def \showURL       {\relax}        \fi
\providecommand\bibfield[2]{#2}
\providecommand\bibinfo[2]{#2}
\providecommand\natexlab[1]{#1}
\providecommand\showeprint[2][]{arXiv:#2}

\bibitem[dan({[n.\,d.]})]%
        {dangerousspeech}
 \bibinfo{year}{[n.\,d.]}\natexlab{}.
\newblock \bibinfo{title}{Dangerous Speech Project}.
\newblock
\urldef\tempurl%
\url{https://dangerousspeech.org/}
\showURL{%
\tempurl}


\bibitem[bir(2021)]%
        {birdwatch}
 \bibinfo{year}{2021}\natexlab{}.
\newblock \bibinfo{title}{Introducing Birdwatch, a community-based approach to misinformation}.
\newblock
\urldef\tempurl%
\url{https://blog.twitter.com/en_us/topics/product/2021/introducing-birdwatch-a-community-based-approach-to-misinformation}
\showURL{%
Retrieved Jan 21, 2021 from \tempurl}


\bibitem[Abas and Khan(2014)]%
        {abas2014carbon}
\bibfield{author}{\bibinfo{person}{N Abas} {and} \bibinfo{person}{N Khan}.} \bibinfo{year}{2014}\natexlab{}.
\newblock \showarticletitle{Carbon conundrum, climate change, CO2 capture and consumptions}.
\newblock \bibinfo{journal}{\emph{Journal of CO2 Utilization}}  \bibinfo{volume}{8} (\bibinfo{year}{2014}), \bibinfo{pages}{39--48}.
\newblock


\bibitem[Ackerman(2000)]%
        {ackerman2000intellectual}
\bibfield{author}{\bibinfo{person}{Mark~S Ackerman}.} \bibinfo{year}{2000}\natexlab{}.
\newblock \showarticletitle{The intellectual challenge of CSCW: the gap between social requirements and technical feasibility}.
\newblock \bibinfo{journal}{\emph{Human--Computer Interaction}} \bibinfo{volume}{15}, \bibinfo{number}{2-3} (\bibinfo{year}{2000}), \bibinfo{pages}{179--203}.
\newblock


\bibitem[Aduragba et~al\mbox{.}(2021)]%
        {aduragba2021detecting}
\bibfield{author}{\bibinfo{person}{Olanrewaju~Tahir Aduragba}, \bibinfo{person}{Jialin Yu}, \bibinfo{person}{Alexandra~I Cristea}, {and} \bibinfo{person}{Lei Shi}.} \bibinfo{year}{2021}\natexlab{}.
\newblock \showarticletitle{Detecting fine-grained emotions on social media during major disease outbreaks: health and well-being before and during the COVID-19 pandemic}. In \bibinfo{booktitle}{\emph{AMIA annual symposium proceedings}}, Vol.~\bibinfo{volume}{2021}. American Medical Informatics Association, \bibinfo{pages}{187}.
\newblock


\bibitem[Aghajari et~al\mbox{.}(2023)]%
        {aghajari2023reviewing}
\bibfield{author}{\bibinfo{person}{Zhila Aghajari}, \bibinfo{person}{Eric~PS Baumer}, {and} \bibinfo{person}{Dominic DiFranzo}.} \bibinfo{year}{2023}\natexlab{}.
\newblock \showarticletitle{Reviewing Interventions to Address Misinformation: The Need to Expand Our Vision Beyond an Individualistic Focus}.
\newblock \bibinfo{journal}{\emph{Proceedings of the ACM on Human-Computer Interaction}} \bibinfo{volume}{7}, \bibinfo{number}{CSCW1} (\bibinfo{year}{2023}), \bibinfo{pages}{1--34}.
\newblock


\bibitem[Albarqouni et~al\mbox{.}(2018)]%
        {albarqouni2018evidence}
\bibfield{author}{\bibinfo{person}{Loai Albarqouni}, \bibinfo{person}{Tammy Hoffmann}, {and} \bibinfo{person}{Paul Glasziou}.} \bibinfo{year}{2018}\natexlab{}.
\newblock \showarticletitle{Evidence-based practice educational intervention studies: a systematic review of what is taught and how it is measured}.
\newblock \bibinfo{journal}{\emph{BMC medical education}} \bibinfo{volume}{18}, \bibinfo{number}{1} (\bibinfo{year}{2018}), \bibinfo{pages}{1--8}.
\newblock


\bibitem[Amazeen et~al\mbox{.}(2022)]%
        {amazeen2022cutting}
\bibfield{author}{\bibinfo{person}{Michelle~A Amazeen}, \bibinfo{person}{Arunima Krishna}, {and} \bibinfo{person}{Rob Eschmann}.} \bibinfo{year}{2022}\natexlab{}.
\newblock \showarticletitle{Cutting the bunk: Comparing the solo and aggregate effects of prebunking and debunking COVID-19 vaccine misinformation}.
\newblock \bibinfo{journal}{\emph{Science Communication}} \bibinfo{volume}{44}, \bibinfo{number}{4} (\bibinfo{year}{2022}), \bibinfo{pages}{387--417}.
\newblock


\bibitem[Anderson et~al\mbox{.}(1980)]%
        {anderson1980perseverance}
\bibfield{author}{\bibinfo{person}{Craig~A Anderson}, \bibinfo{person}{Mark~R Lepper}, {and} \bibinfo{person}{Lee Ross}.} \bibinfo{year}{1980}\natexlab{}.
\newblock \showarticletitle{Perseverance of social theories: The role of explanation in the persistence of discredited information.}
\newblock \bibinfo{journal}{\emph{Journal of personality and social psychology}} \bibinfo{volume}{39}, \bibinfo{number}{6} (\bibinfo{year}{1980}), \bibinfo{pages}{1037}.
\newblock


\bibitem[Baughan et~al\mbox{.}(2022)]%
        {baughan2022shame}
\bibfield{author}{\bibinfo{person}{Amanda Baughan}, \bibinfo{person}{Katherine~Alejandra Cross}, \bibinfo{person}{Elena Khasanova}, {and} \bibinfo{person}{Alexis Hiniker}.} \bibinfo{year}{2022}\natexlab{}.
\newblock \showarticletitle{Shame on Who? Experimentally Reducing Shame During Political Arguments on Twitter}.
\newblock \bibinfo{journal}{\emph{Proceedings of the ACM on Human-Computer Interaction}} \bibinfo{volume}{6}, \bibinfo{number}{CSCW2} (\bibinfo{year}{2022}), \bibinfo{pages}{1--18}.
\newblock


\bibitem[Benegal and Scruggs(2018)]%
        {benegal2018correcting}
\bibfield{author}{\bibinfo{person}{Salil~D Benegal} {and} \bibinfo{person}{Lyle~A Scruggs}.} \bibinfo{year}{2018}\natexlab{}.
\newblock \showarticletitle{Correcting misinformation about climate change: The impact of partisanship in an experimental setting}.
\newblock \bibinfo{journal}{\emph{Climatic change}} \bibinfo{volume}{148}, \bibinfo{number}{1-2} (\bibinfo{year}{2018}), \bibinfo{pages}{61--80}.
\newblock


\bibitem[Berardelli(2020)]%
        {cbs}
\bibfield{author}{\bibinfo{person}{Jeff Berardelli}.} \bibinfo{year}{2020}\natexlab{}.
\newblock \bibinfo{title}{10 common myths about climate change — and what science really says}.
\newblock
\urldef\tempurl%
\url{https://www.cbsnews.com/news/climate-change-myths-what-science-really-says/}
\showURL{%
\tempurl}


\bibitem[Bessi et~al\mbox{.}(2015)]%
        {bessi2015science}
\bibfield{author}{\bibinfo{person}{Alessandro Bessi}, \bibinfo{person}{Mauro Coletto}, \bibinfo{person}{George~Alexandru Davidescu}, \bibinfo{person}{Antonio Scala}, \bibinfo{person}{Guido Caldarelli}, {and} \bibinfo{person}{Walter Quattrociocchi}.} \bibinfo{year}{2015}\natexlab{}.
\newblock \showarticletitle{Science vs conspiracy: Collective narratives in the age of misinformation}.
\newblock \bibinfo{journal}{\emph{PloS one}} \bibinfo{volume}{10}, \bibinfo{number}{2} (\bibinfo{year}{2015}), \bibinfo{pages}{e0118093}.
\newblock


\bibitem[Brown(1982)]%
        {brown1982attention}
\bibfield{author}{\bibinfo{person}{William~R Brown}.} \bibinfo{year}{1982}\natexlab{}.
\newblock \showarticletitle{Attention and the rhetoric of social intervention}.
\newblock \bibinfo{journal}{\emph{Quarterly Journal of Speech}} \bibinfo{volume}{68}, \bibinfo{number}{1} (\bibinfo{year}{1982}), \bibinfo{pages}{17--27}.
\newblock


\bibitem[Byford(2011)]%
        {byford2011conspiracy}
\bibfield{author}{\bibinfo{person}{Jovan Byford}.} \bibinfo{year}{2011}\natexlab{}.
\newblock \bibinfo{booktitle}{\emph{Conspiracy theories: A critical introduction}}.
\newblock \bibinfo{publisher}{Springer}.
\newblock


\bibitem[Cazenave and Remy(2011)]%
        {cazenave2011sea}
\bibfield{author}{\bibinfo{person}{Anny Cazenave} {and} \bibinfo{person}{Frederique Remy}.} \bibinfo{year}{2011}\natexlab{}.
\newblock \showarticletitle{Sea level and climate: measurements and causes of changes}.
\newblock \bibinfo{journal}{\emph{Wiley Interdisciplinary Reviews: Climate Change}} \bibinfo{volume}{2}, \bibinfo{number}{5} (\bibinfo{year}{2011}), \bibinfo{pages}{647--662}.
\newblock


\bibitem[Chan et~al\mbox{.}(2017)]%
        {chan2017debunking}
\bibfield{author}{\bibinfo{person}{Man-pui~Sally Chan}, \bibinfo{person}{Christopher~R Jones}, \bibinfo{person}{Kathleen Hall~Jamieson}, {and} \bibinfo{person}{Dolores Albarrac{\'\i}n}.} \bibinfo{year}{2017}\natexlab{}.
\newblock \showarticletitle{Debunking: A meta-analysis of the psychological efficacy of messages countering misinformation}.
\newblock \bibinfo{journal}{\emph{Psychological science}} \bibinfo{volume}{28}, \bibinfo{number}{11} (\bibinfo{year}{2017}), \bibinfo{pages}{1531--1546}.
\newblock


\bibitem[Christian et~al\mbox{.}(2016)]%
        {christian2016single}
\bibfield{author}{\bibinfo{person}{Hans Christian}, \bibinfo{person}{Mikhael~Pramodana Agus}, {and} \bibinfo{person}{Derwin Suhartono}.} \bibinfo{year}{2016}\natexlab{}.
\newblock \showarticletitle{Single document automatic text summarization using term frequency-inverse document frequency (TF-IDF)}.
\newblock \bibinfo{journal}{\emph{ComTech: Computer, Mathematics and Engineering Applications}} \bibinfo{volume}{7}, \bibinfo{number}{4} (\bibinfo{year}{2016}), \bibinfo{pages}{285--294}.
\newblock


\bibitem[Cinelli et~al\mbox{.}(2022)]%
        {cinelli2022conspiracy}
\bibfield{author}{\bibinfo{person}{Matteo Cinelli}, \bibinfo{person}{Gabriele Etta}, \bibinfo{person}{Michele Avalle}, \bibinfo{person}{Alessandro Quattrociocchi}, \bibinfo{person}{Niccol{\`o} Di~Marco}, \bibinfo{person}{Carlo Valensise}, \bibinfo{person}{Alessandro Galeazzi}, {and} \bibinfo{person}{Walter Quattrociocchi}.} \bibinfo{year}{2022}\natexlab{}.
\newblock \showarticletitle{Conspiracy theories and social media platforms}.
\newblock \bibinfo{journal}{\emph{Current Opinion in Psychology}} (\bibinfo{year}{2022}), \bibinfo{pages}{101407}.
\newblock


\bibitem[Communication(2023)]%
        {sixamerica}
\bibfield{author}{\bibinfo{person}{Yale Climate~Change Communication}.} \bibinfo{year}{2023}\natexlab{}.
\newblock \bibinfo{title}{Global Warming's Six Americas}.
\newblock
\urldef\tempurl%
\url{https://climatecommunication.yale.edu/about/projects/global-warmings-six-americas/}
\showURL{%
\tempurl}


\bibitem[Cook(2020)]%
        {cook2020deconstructing}
\bibfield{author}{\bibinfo{person}{John Cook}.} \bibinfo{year}{2020}\natexlab{}.
\newblock \showarticletitle{Deconstructing climate science denial}.
\newblock \bibinfo{journal}{\emph{Research handbook on communicating climate change}} (\bibinfo{year}{2020}), \bibinfo{pages}{62--78}.
\newblock


\bibitem[Corner et~al\mbox{.}(2014)]%
        {corner2014public}
\bibfield{author}{\bibinfo{person}{Adam Corner}, \bibinfo{person}{Ezra Markowitz}, {and} \bibinfo{person}{Nick Pidgeon}.} \bibinfo{year}{2014}\natexlab{}.
\newblock \showarticletitle{Public engagement with climate change: the role of human values}.
\newblock \bibinfo{journal}{\emph{Wiley interdisciplinary reviews: climate change}} \bibinfo{volume}{5}, \bibinfo{number}{3} (\bibinfo{year}{2014}), \bibinfo{pages}{411--422}.
\newblock


\bibitem[Dentith(2019)]%
        {dentith2019conspiracy}
\bibfield{author}{\bibinfo{person}{Matthew~RX Dentith}.} \bibinfo{year}{2019}\natexlab{}.
\newblock \showarticletitle{Conspiracy theories on the basis of the evidence}.
\newblock \bibinfo{journal}{\emph{Synthese}}  \bibinfo{volume}{196} (\bibinfo{year}{2019}), \bibinfo{pages}{2243--2261}.
\newblock


\bibitem[Dentith and Orr(2018)]%
        {dentith2018secrecy}
\bibfield{author}{\bibinfo{person}{Matthew~RX Dentith} {and} \bibinfo{person}{Martin Orr}.} \bibinfo{year}{2018}\natexlab{}.
\newblock \showarticletitle{Secrecy and conspiracy}.
\newblock \bibinfo{journal}{\emph{Episteme}} \bibinfo{volume}{15}, \bibinfo{number}{4} (\bibinfo{year}{2018}), \bibinfo{pages}{433--450}.
\newblock


\bibitem[Douglas and Sutton(2015)]%
        {douglas2015climate}
\bibfield{author}{\bibinfo{person}{Karen~M Douglas} {and} \bibinfo{person}{Robbie~M Sutton}.} \bibinfo{year}{2015}\natexlab{}.
\newblock \showarticletitle{Climate change: Why the conspiracy theories are dangerous}.
\newblock \bibinfo{journal}{\emph{Bulletin of the Atomic Scientists}} \bibinfo{volume}{71}, \bibinfo{number}{2} (\bibinfo{year}{2015}), \bibinfo{pages}{98--106}.
\newblock


\bibitem[Douglas et~al\mbox{.}(2017)]%
        {douglas2017psychology}
\bibfield{author}{\bibinfo{person}{Karen~M Douglas}, \bibinfo{person}{Robbie~M Sutton}, {and} \bibinfo{person}{Aleksandra Cichocka}.} \bibinfo{year}{2017}\natexlab{}.
\newblock \showarticletitle{The psychology of conspiracy theories}.
\newblock \bibinfo{journal}{\emph{Current directions in psychological science}} \bibinfo{volume}{26}, \bibinfo{number}{6} (\bibinfo{year}{2017}), \bibinfo{pages}{538--542}.
\newblock


\bibitem[Douglas et~al\mbox{.}(2015)]%
        {douglas2015social}
\bibfield{author}{\bibinfo{person}{Karen~M Douglas}, \bibinfo{person}{Robbie~M Sutton}, \bibinfo{person}{Daniel Jolley}, {and} \bibinfo{person}{Michael~J Wood}.} \bibinfo{year}{2015}\natexlab{}.
\newblock \showarticletitle{The social, political, environmental, and health-related consequences of conspiracy theories}.
\newblock \bibinfo{journal}{\emph{The psychology of conspiracy}}  \bibinfo{volume}{56} (\bibinfo{year}{2015}), \bibinfo{pages}{183--200}.
\newblock


\bibitem[Douglas et~al\mbox{.}(2019)]%
        {douglas2019understanding}
\bibfield{author}{\bibinfo{person}{Karen~M Douglas}, \bibinfo{person}{Joseph~E Uscinski}, \bibinfo{person}{Robbie~M Sutton}, \bibinfo{person}{Aleksandra Cichocka}, \bibinfo{person}{Turkay Nefes}, \bibinfo{person}{Chee~Siang Ang}, {and} \bibinfo{person}{Farzin Deravi}.} \bibinfo{year}{2019}\natexlab{}.
\newblock \showarticletitle{Understanding conspiracy theories}.
\newblock \bibinfo{journal}{\emph{Political psychology}}  \bibinfo{volume}{40} (\bibinfo{year}{2019}), \bibinfo{pages}{3--35}.
\newblock


\bibitem[Drolsbach and Pr{\"o}llochs(2023)]%
        {drolsbach2023diffusion}
\bibfield{author}{\bibinfo{person}{Chiara~Patricia Drolsbach} {and} \bibinfo{person}{Nicolas Pr{\"o}llochs}.} \bibinfo{year}{2023}\natexlab{}.
\newblock \showarticletitle{Diffusion of community fact-checked misinformation on Twitter}.
\newblock \bibinfo{journal}{\emph{Proceedings of the ACM on Human-Computer Interaction}} \bibinfo{volume}{7}, \bibinfo{number}{CSCW2} (\bibinfo{year}{2023}), \bibinfo{pages}{1--22}.
\newblock


\bibitem[Dunlap et~al\mbox{.}(2011)]%
        {dunlap2011organized}
\bibfield{author}{\bibinfo{person}{Riley~E Dunlap}, \bibinfo{person}{Aaron~M McCright}, {et~al\mbox{.}}} \bibinfo{year}{2011}\natexlab{}.
\newblock \showarticletitle{Organized climate change denial}.
\newblock \bibinfo{journal}{\emph{The Oxford handbook of climate change and society}}  \bibinfo{volume}{1} (\bibinfo{year}{2011}), \bibinfo{pages}{144--160}.
\newblock


\bibitem[Dwyer(2007)]%
        {dwyer2007task}
\bibfield{author}{\bibinfo{person}{Catherine Dwyer}.} \bibinfo{year}{2007}\natexlab{}.
\newblock \showarticletitle{Task Technology Fit, the social technical gap and social networking sites}.
\newblock \bibinfo{journal}{\emph{AMCIS 2007 Proceedings}} (\bibinfo{year}{2007}), \bibinfo{pages}{374}.
\newblock


\bibitem[Eisenstein et~al\mbox{.}(2011)]%
        {eisenstein2011sparse}
\bibfield{author}{\bibinfo{person}{Jacob Eisenstein}, \bibinfo{person}{Amr Ahmed}, {and} \bibinfo{person}{Eric~P Xing}.} \bibinfo{year}{2011}\natexlab{}.
\newblock \showarticletitle{Sparse additive generative models of text}. In \bibinfo{booktitle}{\emph{Proceedings of the 28th international conference on machine learning (ICML-11)}}. \bibinfo{pages}{1041--1048}.
\newblock


\bibitem[Engel et~al\mbox{.}(2023)]%
        {engel2023learning}
\bibfield{author}{\bibinfo{person}{Kristen Engel}, \bibinfo{person}{Shruti Phadke}, {and} \bibinfo{person}{Tanushree Mitra}.} \bibinfo{year}{2023}\natexlab{}.
\newblock \showarticletitle{Learning from the Ex-Believers: Individuals' Journeys In and Out of Conspiracy Theories Online}.
\newblock \bibinfo{journal}{\emph{Proceedings of the ACM on Human-Computer Interaction}} \bibinfo{volume}{7}, \bibinfo{number}{CSCW2} (\bibinfo{year}{2023}), \bibinfo{pages}{1--37}.
\newblock


\bibitem[Etkins and Epstein(1982)]%
        {etkins1982rise}
\bibfield{author}{\bibinfo{person}{Robert Etkins} {and} \bibinfo{person}{Edward~S Epstein}.} \bibinfo{year}{1982}\natexlab{}.
\newblock \showarticletitle{The rise of global mean sea level as an indication of climate change}.
\newblock \bibinfo{journal}{\emph{Science}} \bibinfo{volume}{215}, \bibinfo{number}{4530} (\bibinfo{year}{1982}), \bibinfo{pages}{287--289}.
\newblock


\bibitem[Flynn et~al\mbox{.}(2017)]%
        {flynn2017nature}
\bibfield{author}{\bibinfo{person}{Daniel~J Flynn}, \bibinfo{person}{Brendan Nyhan}, {and} \bibinfo{person}{Jason Reifler}.} \bibinfo{year}{2017}\natexlab{}.
\newblock \showarticletitle{The nature and origins of misperceptions: Understanding false and unsupported beliefs about politics}.
\newblock \bibinfo{journal}{\emph{Political Psychology}}  \bibinfo{volume}{38} (\bibinfo{year}{2017}), \bibinfo{pages}{127--150}.
\newblock


\bibitem[Freiling et~al\mbox{.}(2023)]%
        {freiling2023science}
\bibfield{author}{\bibinfo{person}{Isabelle Freiling}, \bibinfo{person}{Nicole~M Krause}, {and} \bibinfo{person}{Dietram~A Scheufele}.} \bibinfo{year}{2023}\natexlab{}.
\newblock \showarticletitle{Science and ethics of “curing” misinformation}.
\newblock \bibinfo{journal}{\emph{AMA journal of ethics}} \bibinfo{volume}{25}, \bibinfo{number}{3} (\bibinfo{year}{2023}), \bibinfo{pages}{228--237}.
\newblock


\bibitem[Gargiulo(2006)]%
        {gargiulo2006power}
\bibfield{author}{\bibinfo{person}{Terrence~L Gargiulo}.} \bibinfo{year}{2006}\natexlab{}.
\newblock \showarticletitle{Power of stories}.
\newblock \bibinfo{journal}{\emph{The journal for quality and participation}} \bibinfo{volume}{29}, \bibinfo{number}{1} (\bibinfo{year}{2006}), \bibinfo{pages}{4}.
\newblock


\bibitem[Garrett and Weeks(2013)]%
        {garrett2013promise}
\bibfield{author}{\bibinfo{person}{R~Kelly Garrett} {and} \bibinfo{person}{Brian~E Weeks}.} \bibinfo{year}{2013}\natexlab{}.
\newblock \showarticletitle{The promise and peril of real-time corrections to political misperceptions}. In \bibinfo{booktitle}{\emph{Proceedings of the 2013 conference on Computer supported cooperative work}}. \bibinfo{pages}{1047--1058}.
\newblock


\bibitem[Girardin(2007)]%
        {girardin2007towards}
\bibfield{author}{\bibinfo{person}{Fabien Girardin}.} \bibinfo{year}{2007}\natexlab{}.
\newblock \emph{\bibinfo{title}{Towards Reducing the Social-Technical Gap in Location-Aware Computing}}.
\newblock \bibinfo{thesistype}{Ph.\,D. Dissertation}. \bibinfo{school}{Ph. D. Dissertation. Citeseer}.
\newblock


\bibitem[Goreis et~al\mbox{.}(2020)]%
        {goreis2020social}
\bibfield{author}{\bibinfo{person}{Andreas Goreis}, \bibinfo{person}{Oswald~D Kothgassner}, {et~al\mbox{.}}} \bibinfo{year}{2020}\natexlab{}.
\newblock \showarticletitle{Social media as vehicle for conspiracy beliefs on COVID-19}.
\newblock \bibinfo{journal}{\emph{Digital Psychology}} \bibinfo{volume}{1}, \bibinfo{number}{2} (\bibinfo{year}{2020}), \bibinfo{pages}{36--39}.
\newblock


\bibitem[Graham et~al\mbox{.}(1990)]%
        {graham1990increasing}
\bibfield{author}{\bibinfo{person}{Robin~Lambert Graham}, \bibinfo{person}{Monica~G Turner}, {and} \bibinfo{person}{Virginia~H Dale}.} \bibinfo{year}{1990}\natexlab{}.
\newblock \showarticletitle{How increasing CO2 and climate change affect forests}.
\newblock \bibinfo{journal}{\emph{BioScience}} \bibinfo{volume}{40}, \bibinfo{number}{8} (\bibinfo{year}{1990}), \bibinfo{pages}{575--587}.
\newblock


\bibitem[Hangartner et~al\mbox{.}(2021)]%
        {hangartner2021empathy}
\bibfield{author}{\bibinfo{person}{Dominik Hangartner}, \bibinfo{person}{Gloria Gennaro}, \bibinfo{person}{Sary Alasiri}, \bibinfo{person}{Nicholas Bahrich}, \bibinfo{person}{Alexandra Bornhoft}, \bibinfo{person}{Joseph Boucher}, \bibinfo{person}{Buket~Buse Demirci}, \bibinfo{person}{Laurenz Derksen}, \bibinfo{person}{Aldo Hall}, \bibinfo{person}{Matthias Jochum}, {et~al\mbox{.}}} \bibinfo{year}{2021}\natexlab{}.
\newblock \showarticletitle{Empathy-based counterspeech can reduce racist hate speech in a social media field experiment}.
\newblock \bibinfo{journal}{\emph{Proceedings of the National Academy of Sciences}} \bibinfo{volume}{118}, \bibinfo{number}{50} (\bibinfo{year}{2021}), \bibinfo{pages}{e2116310118}.
\newblock


\bibitem[Heath et~al\mbox{.}(2001)]%
        {heath2001emotional}
\bibfield{author}{\bibinfo{person}{Chip Heath}, \bibinfo{person}{Chris Bell}, {and} \bibinfo{person}{Emily Sternberg}.} \bibinfo{year}{2001}\natexlab{}.
\newblock \showarticletitle{Emotional selection in memes: the case of urban legends.}
\newblock \bibinfo{journal}{\emph{Journal of personality and social psychology}} \bibinfo{volume}{81}, \bibinfo{number}{6} (\bibinfo{year}{2001}), \bibinfo{pages}{1028}.
\newblock


\bibitem[Help({[n.\,d.]})]%
        {karma}
\bibfield{author}{\bibinfo{person}{Reddit Help}.} \bibinfo{year}{[n.\,d.]}\natexlab{}.
\newblock \bibinfo{title}{What is karma?}
\newblock
\urldef\tempurl%
\url{https://support.reddithelp.com/hc/en-us/articles/204511829-What-is-karma-}
\showURL{%
\tempurl}


\bibitem[Hofstadter(2012)]%
        {hofstadter2012paranoid}
\bibfield{author}{\bibinfo{person}{Richard Hofstadter}.} \bibinfo{year}{2012}\natexlab{}.
\newblock \bibinfo{booktitle}{\emph{The paranoid style in American politics}}.
\newblock \bibinfo{publisher}{Vintage}.
\newblock


\bibitem[Hossain(2023)]%
        {hossain2023sun}
\bibfield{author}{\bibinfo{person}{Eklas Hossain}.} \bibinfo{year}{2023}\natexlab{}.
\newblock \bibinfo{booktitle}{\emph{The sun, energy, and climate change}}.
\newblock \bibinfo{publisher}{Springer}.
\newblock


\bibitem[Jary(1998)]%
        {jary1998relevance}
\bibfield{author}{\bibinfo{person}{Mark Jary}.} \bibinfo{year}{1998}\natexlab{}.
\newblock \showarticletitle{Relevance theory and the communication of politeness}.
\newblock \bibinfo{journal}{\emph{Journal of pragmatics}} \bibinfo{volume}{30}, \bibinfo{number}{1} (\bibinfo{year}{1998}), \bibinfo{pages}{1--19}.
\newblock


\bibitem[Kahan(2013)]%
        {kahan2013ideology}
\bibfield{author}{\bibinfo{person}{Dan~M Kahan}.} \bibinfo{year}{2013}\natexlab{}.
\newblock \showarticletitle{Ideology, motivated reasoning, and cognitive reflection}.
\newblock \bibinfo{journal}{\emph{Judgment and Decision making}} \bibinfo{volume}{8}, \bibinfo{number}{4} (\bibinfo{year}{2013}), \bibinfo{pages}{407--424}.
\newblock


\bibitem[Kahne and Bowyer(2017)]%
        {kahne2017educating}
\bibfield{author}{\bibinfo{person}{Joseph Kahne} {and} \bibinfo{person}{Benjamin Bowyer}.} \bibinfo{year}{2017}\natexlab{}.
\newblock \showarticletitle{Educating for democracy in a partisan age: Confronting the challenges of motivated reasoning and misinformation}.
\newblock \bibinfo{journal}{\emph{American educational research journal}} \bibinfo{volume}{54}, \bibinfo{number}{1} (\bibinfo{year}{2017}), \bibinfo{pages}{3--34}.
\newblock


\bibitem[Kegeles et~al\mbox{.}(2012)]%
        {kegeles2012intervention}
\bibfield{author}{\bibinfo{person}{Susan~M Kegeles}, \bibinfo{person}{Gregory Rebchook}, \bibinfo{person}{Lance Pollack}, \bibinfo{person}{David Huebner}, \bibinfo{person}{Scott Tebbetts}, \bibinfo{person}{John Hamiga}, \bibinfo{person}{David Sweeney}, {and} \bibinfo{person}{Benjamin Zovod}.} \bibinfo{year}{2012}\natexlab{}.
\newblock \showarticletitle{An intervention to help community-based organizations implement an evidence-based HIV prevention intervention: The mpowerment project technology exchange system}.
\newblock \bibinfo{journal}{\emph{American Journal of Community Psychology}} \bibinfo{volume}{49}, \bibinfo{number}{1-2} (\bibinfo{year}{2012}), \bibinfo{pages}{182--198}.
\newblock


\bibitem[Kim and Vossen(2021)]%
        {kim2021emoberta}
\bibfield{author}{\bibinfo{person}{Taewoon Kim} {and} \bibinfo{person}{Piek Vossen}.} \bibinfo{year}{2021}\natexlab{}.
\newblock \showarticletitle{Emoberta: Speaker-aware emotion recognition in conversation with roberta}.
\newblock \bibinfo{journal}{\emph{arXiv preprint arXiv:2108.12009}} (\bibinfo{year}{2021}).
\newblock


\bibitem[Klein et~al\mbox{.}(2019)]%
        {klein2019pathways}
\bibfield{author}{\bibinfo{person}{Colin Klein}, \bibinfo{person}{Peter Clutton}, {and} \bibinfo{person}{Adam~G Dunn}.} \bibinfo{year}{2019}\natexlab{}.
\newblock \showarticletitle{Pathways to conspiracy: The social and linguistic precursors of involvement in Reddit’s conspiracy theory forum}.
\newblock \bibinfo{journal}{\emph{PloS one}} \bibinfo{volume}{14}, \bibinfo{number}{11} (\bibinfo{year}{2019}), \bibinfo{pages}{e0225098}.
\newblock


\bibitem[Koo et~al\mbox{.}(2021)]%
        {koo2021motivates}
\bibfield{author}{\bibinfo{person}{Alex Zhi-Xiong Koo}, \bibinfo{person}{Min-Hsin Su}, \bibinfo{person}{SangWon Lee}, \bibinfo{person}{So-Yun Ahn}, {and} \bibinfo{person}{Hernando Rojas}.} \bibinfo{year}{2021}\natexlab{}.
\newblock \showarticletitle{What motivates people to correct misinformation? Examining the effects of third-person perceptions and perceived norms}.
\newblock \bibinfo{journal}{\emph{Journal of Broadcasting \& Electronic Media}} \bibinfo{volume}{65}, \bibinfo{number}{1} (\bibinfo{year}{2021}), \bibinfo{pages}{111--134}.
\newblock


\bibitem[Kratochwill and Shernoff(2003)]%
        {kratochwill2003evidence}
\bibfield{author}{\bibinfo{person}{Thomas~R Kratochwill} {and} \bibinfo{person}{Elisa~Steele Shernoff}.} \bibinfo{year}{2003}\natexlab{}.
\newblock \showarticletitle{Evidence-based practice: Promoting evidence-based interventions in school psychology.}
\newblock \bibinfo{journal}{\emph{School Psychology Quarterly}} \bibinfo{volume}{18}, \bibinfo{number}{4} (\bibinfo{year}{2003}), \bibinfo{pages}{389}.
\newblock


\bibitem[Kunda(1990)]%
        {kunda1990case}
\bibfield{author}{\bibinfo{person}{Ziva Kunda}.} \bibinfo{year}{1990}\natexlab{}.
\newblock \showarticletitle{The case for motivated reasoning.}
\newblock \bibinfo{journal}{\emph{Psychological bulletin}} \bibinfo{volume}{108}, \bibinfo{number}{3} (\bibinfo{year}{1990}), \bibinfo{pages}{480}.
\newblock


\bibitem[Ku{\.z}elewska and Tomaszuk(2022)]%
        {kuzelewska2022rise}
\bibfield{author}{\bibinfo{person}{El{\.z}bieta Ku{\.z}elewska} {and} \bibinfo{person}{Mariusz Tomaszuk}.} \bibinfo{year}{2022}\natexlab{}.
\newblock \showarticletitle{Rise of conspiracy theories in the pandemic times}.
\newblock \bibinfo{journal}{\emph{International Journal for the Semiotics of Law-Revue internationale de S{\'e}miotique juridique}} \bibinfo{volume}{35}, \bibinfo{number}{6} (\bibinfo{year}{2022}), \bibinfo{pages}{2373--2389}.
\newblock


\bibitem[Lewandowsky et~al\mbox{.}(2020)]%
        {lewandowsky2020debunking}
\bibfield{author}{\bibinfo{person}{Stephan Lewandowsky}, \bibinfo{person}{John Cook}, \bibinfo{person}{Ullrich Ecker}, \bibinfo{person}{Dolores Albarrac{\'\i}n}, \bibinfo{person}{Panayiota Kendeou}, \bibinfo{person}{Eryn~J Newman}, \bibinfo{person}{Gordon Pennycook}, \bibinfo{person}{Ethan Porter}, \bibinfo{person}{David~G Rand}, \bibinfo{person}{David~N Rapp}, {et~al\mbox{.}}} \bibinfo{year}{2020}\natexlab{}.
\newblock \showarticletitle{The debunking handbook 2020}.
\newblock  (\bibinfo{year}{2020}).
\newblock


\bibitem[Lewandowsky et~al\mbox{.}(2017)]%
        {lewandowsky2017beyond}
\bibfield{author}{\bibinfo{person}{Stephan Lewandowsky}, \bibinfo{person}{Ullrich~KH Ecker}, {and} \bibinfo{person}{John Cook}.} \bibinfo{year}{2017}\natexlab{}.
\newblock \showarticletitle{Beyond misinformation: Understanding and coping with the “post-truth” era}.
\newblock \bibinfo{journal}{\emph{Journal of applied research in memory and cognition}} \bibinfo{volume}{6}, \bibinfo{number}{4} (\bibinfo{year}{2017}), \bibinfo{pages}{353--369}.
\newblock


\bibitem[Lewandowsky et~al\mbox{.}(2015)]%
        {lewandowsky2015seepage}
\bibfield{author}{\bibinfo{person}{Stephan Lewandowsky}, \bibinfo{person}{Naomi Oreskes}, \bibinfo{person}{James~S Risbey}, \bibinfo{person}{Ben~R Newell}, {and} \bibinfo{person}{Michael Smithson}.} \bibinfo{year}{2015}\natexlab{}.
\newblock \showarticletitle{Seepage: Climate change denial and its effect on the scientific community}.
\newblock \bibinfo{journal}{\emph{Global Environmental Change}}  \bibinfo{volume}{33} (\bibinfo{year}{2015}), \bibinfo{pages}{1--13}.
\newblock


\bibitem[Lindzen(1997)]%
        {lindzen1997can}
\bibfield{author}{\bibinfo{person}{Richard~S Lindzen}.} \bibinfo{year}{1997}\natexlab{}.
\newblock \showarticletitle{Can increasing carbon dioxide cause climate change?}
\newblock \bibinfo{journal}{\emph{Proceedings of the National Academy of Sciences}} \bibinfo{volume}{94}, \bibinfo{number}{16} (\bibinfo{year}{1997}), \bibinfo{pages}{8335--8342}.
\newblock


\bibitem[Lodge and Taber(2013)]%
        {lodge2013rationalizing}
\bibfield{author}{\bibinfo{person}{Milton Lodge} {and} \bibinfo{person}{Charles~S Taber}.} \bibinfo{year}{2013}\natexlab{}.
\newblock \bibinfo{booktitle}{\emph{The rationalizing voter}}.
\newblock \bibinfo{publisher}{Cambridge University Press}.
\newblock


\bibitem[Lynn(2018)]%
        {lynn2018communicating}
\bibfield{author}{\bibinfo{person}{Jonathan Lynn}.} \bibinfo{year}{2018}\natexlab{}.
\newblock \showarticletitle{Communicating the IPCC: challenges and opportunities}.
\newblock \bibinfo{journal}{\emph{Handbook of Climate Change Communication: Vol. 3: Case Studies in Climate Change Communication}} (\bibinfo{year}{2018}), \bibinfo{pages}{131--143}.
\newblock


\bibitem[Mackenzie and Wallace(2011)]%
        {mackenzie2011communication}
\bibfield{author}{\bibinfo{person}{Lauren Mackenzie} {and} \bibinfo{person}{Megan Wallace}.} \bibinfo{year}{2011}\natexlab{}.
\newblock \showarticletitle{The communication of respect as a significant dimension of cross-cultural communication competence}.
\newblock \bibinfo{journal}{\emph{Cross-Cultural Communication}} \bibinfo{volume}{7}, \bibinfo{number}{3} (\bibinfo{year}{2011}), \bibinfo{pages}{10--18}.
\newblock


\bibitem[Mimura(2013)]%
        {mimura2013sea}
\bibfield{author}{\bibinfo{person}{Nobuo Mimura}.} \bibinfo{year}{2013}\natexlab{}.
\newblock \showarticletitle{Sea-level rise caused by climate change and its implications for society}.
\newblock \bibinfo{journal}{\emph{Proceedings of the Japan Academy, Series B}} \bibinfo{volume}{89}, \bibinfo{number}{7} (\bibinfo{year}{2013}), \bibinfo{pages}{281--301}.
\newblock


\bibitem[Montzka et~al\mbox{.}(2011)]%
        {montzka2011non}
\bibfield{author}{\bibinfo{person}{Stephen~A Montzka}, \bibinfo{person}{Edward~J Dlugokencky}, {and} \bibinfo{person}{James~H Butler}.} \bibinfo{year}{2011}\natexlab{}.
\newblock \showarticletitle{Non-CO2 greenhouse gases and climate change}.
\newblock \bibinfo{journal}{\emph{Nature}} \bibinfo{volume}{476}, \bibinfo{number}{7358} (\bibinfo{year}{2011}), \bibinfo{pages}{43--50}.
\newblock


\bibitem[Moravec et~al\mbox{.}(2018)]%
        {moravec2018fake}
\bibfield{author}{\bibinfo{person}{Patricia Moravec}, \bibinfo{person}{Randall Minas}, {and} \bibinfo{person}{Alan~R Dennis}.} \bibinfo{year}{2018}\natexlab{}.
\newblock \showarticletitle{Fake news on social media: People believe what they want to believe when it makes no sense at all}.
\newblock \bibinfo{journal}{\emph{Kelley School of Business research paper}} \bibinfo{number}{18-87} (\bibinfo{year}{2018}).
\newblock


\bibitem[Moyer(2019)]%
        {moyer2019we}
\bibfield{author}{\bibinfo{person}{Melinda~Wenner Moyer}.} \bibinfo{year}{2019}\natexlab{}.
\newblock \showarticletitle{Why we believe conspiracy theories}.
\newblock \bibinfo{journal}{\emph{Scientific American}} \bibinfo{volume}{320}, \bibinfo{number}{3} (\bibinfo{year}{2019}), \bibinfo{pages}{58--63}.
\newblock


\bibitem[Munger(2017)]%
        {munger2017tweetment}
\bibfield{author}{\bibinfo{person}{Kevin Munger}.} \bibinfo{year}{2017}\natexlab{}.
\newblock \showarticletitle{Tweetment effects on the tweeted: Experimentally reducing racist harassment}.
\newblock \bibinfo{journal}{\emph{Political Behavior}}  \bibinfo{volume}{39} (\bibinfo{year}{2017}), \bibinfo{pages}{629--649}.
\newblock


\bibitem[Nera et~al\mbox{.}(2022)]%
        {nera2022looking}
\bibfield{author}{\bibinfo{person}{Kenzo Nera}, \bibinfo{person}{Youri~L Mora}, \bibinfo{person}{Pit Klein}, \bibinfo{person}{Antoine Roblain}, \bibinfo{person}{Pascaline Van~Oost}, \bibinfo{person}{Julie Terache}, {and} \bibinfo{person}{Olivier Klein}.} \bibinfo{year}{2022}\natexlab{}.
\newblock \showarticletitle{Looking for ties with secret agendas during the pandemic: Conspiracy mentality is associated with reduced trust in political, medical, and scientific institutions--but not in medical personnel}.
\newblock \bibinfo{journal}{\emph{Psychologica Belgica}} \bibinfo{volume}{62}, \bibinfo{number}{1} (\bibinfo{year}{2022}), \bibinfo{pages}{193}.
\newblock


\bibitem[Nickerson(1998)]%
        {nickerson1998confirmation}
\bibfield{author}{\bibinfo{person}{Raymond~S Nickerson}.} \bibinfo{year}{1998}\natexlab{}.
\newblock \showarticletitle{Confirmation bias: A ubiquitous phenomenon in many guises}.
\newblock \bibinfo{journal}{\emph{Review of general psychology}} \bibinfo{volume}{2}, \bibinfo{number}{2} (\bibinfo{year}{1998}), \bibinfo{pages}{175--220}.
\newblock


\bibitem[Nyhan and Reifler(2010)]%
        {nyhan2010corrections}
\bibfield{author}{\bibinfo{person}{Brendan Nyhan} {and} \bibinfo{person}{Jason Reifler}.} \bibinfo{year}{2010}\natexlab{}.
\newblock \showarticletitle{When corrections fail: The persistence of political misperceptions}.
\newblock \bibinfo{journal}{\emph{Political Behavior}} \bibinfo{volume}{32}, \bibinfo{number}{2} (\bibinfo{year}{2010}), \bibinfo{pages}{303--330}.
\newblock


\bibitem[Painter and Ashe(2012)]%
        {painter2012cross}
\bibfield{author}{\bibinfo{person}{James Painter} {and} \bibinfo{person}{Teresa Ashe}.} \bibinfo{year}{2012}\natexlab{}.
\newblock \showarticletitle{Cross-national comparison of the presence of climate scepticism in the print media in six countries, 2007--10}.
\newblock \bibinfo{journal}{\emph{Environmental research letters}} \bibinfo{volume}{7}, \bibinfo{number}{4} (\bibinfo{year}{2012}), \bibinfo{pages}{044005}.
\newblock


\bibitem[Pennycook et~al\mbox{.}(2020)]%
        {pennycook2020fighting}
\bibfield{author}{\bibinfo{person}{Gordon Pennycook}, \bibinfo{person}{Jonathon McPhetres}, \bibinfo{person}{Yunhao Zhang}, \bibinfo{person}{Jackson~G Lu}, {and} \bibinfo{person}{David~G Rand}.} \bibinfo{year}{2020}\natexlab{}.
\newblock \showarticletitle{Fighting COVID-19 misinformation on social media: Experimental evidence for a scalable accuracy-nudge intervention}.
\newblock \bibinfo{journal}{\emph{Psychological science}} \bibinfo{volume}{31}, \bibinfo{number}{7} (\bibinfo{year}{2020}), \bibinfo{pages}{770--780}.
\newblock


\bibitem[Phadke et~al\mbox{.}(2021)]%
        {phadke2021characterizing}
\bibfield{author}{\bibinfo{person}{Shruti Phadke}, \bibinfo{person}{Mattia Samory}, {and} \bibinfo{person}{Tanushree Mitra}.} \bibinfo{year}{2021}\natexlab{}.
\newblock \showarticletitle{Characterizing social imaginaries and self-disclosures of dissonance in online conspiracy discussion communities}.
\newblock \bibinfo{journal}{\emph{Proceedings of the ACM on Human-Computer Interaction}} \bibinfo{volume}{5}, \bibinfo{number}{CSCW2} (\bibinfo{year}{2021}), \bibinfo{pages}{1--35}.
\newblock


\bibitem[Phadke et~al\mbox{.}(2022)]%
        {phadke2022pathways}
\bibfield{author}{\bibinfo{person}{Shruti Phadke}, \bibinfo{person}{Mattia Samory}, {and} \bibinfo{person}{Tanushree Mitra}.} \bibinfo{year}{2022}\natexlab{}.
\newblock \showarticletitle{Pathways through conspiracy: the evolution of conspiracy radicalization through engagement in online conspiracy discussions}. In \bibinfo{booktitle}{\emph{Proceedings of the International AAAI Conference on Web and Social Media}}, Vol.~\bibinfo{volume}{16}. \bibinfo{pages}{770--781}.
\newblock


\bibitem[Pr{\"o}llochs and Feuerriegel(2023)]%
        {prollochs2023mechanisms}
\bibfield{author}{\bibinfo{person}{Nicolas Pr{\"o}llochs} {and} \bibinfo{person}{Stefan Feuerriegel}.} \bibinfo{year}{2023}\natexlab{}.
\newblock \showarticletitle{Mechanisms of true and false rumor sharing in social media: Collective intelligence or herd behavior?}
\newblock \bibinfo{journal}{\emph{Proceedings of the ACM on Human-Computer Interaction}} \bibinfo{volume}{7}, \bibinfo{number}{CSCW2} (\bibinfo{year}{2023}), \bibinfo{pages}{1--38}.
\newblock


\bibitem[Prooijen(2018)]%
        {prooijen2018psychology}
\bibfield{author}{\bibinfo{person}{Jan-Willem Prooijen}.} \bibinfo{year}{2018}\natexlab{}.
\newblock \bibinfo{booktitle}{\emph{The psychology of conspiracy theories}}.
\newblock \bibinfo{publisher}{Routledge}.
\newblock


\bibitem[Rocha-Silva et~al\mbox{.}(2023)]%
        {rocha2023passive}
\bibfield{author}{\bibinfo{person}{Tiago Rocha-Silva}, \bibinfo{person}{Concei{\c{c}}{\~a}o Nogueira}, {and} \bibinfo{person}{Liliana Rodrigues}.} \bibinfo{year}{2023}\natexlab{}.
\newblock \showarticletitle{Passive data collection on Reddit: a practical approach}.
\newblock \bibinfo{journal}{\emph{Research Ethics}} (\bibinfo{year}{2023}), \bibinfo{pages}{17470161231210542}.
\newblock


\bibitem[Roozenbeek et~al\mbox{.}(2022)]%
        {roozenbeek2022psychological}
\bibfield{author}{\bibinfo{person}{Jon Roozenbeek}, \bibinfo{person}{Sander Van Der~Linden}, \bibinfo{person}{Beth Goldberg}, \bibinfo{person}{Steve Rathje}, {and} \bibinfo{person}{Stephan Lewandowsky}.} \bibinfo{year}{2022}\natexlab{}.
\newblock \showarticletitle{Psychological inoculation improves resilience against misinformation on social media}.
\newblock \bibinfo{journal}{\emph{Science Advances}} \bibinfo{volume}{8}, \bibinfo{number}{34} (\bibinfo{year}{2022}), \bibinfo{pages}{eabo6254}.
\newblock


\bibitem[Ryabova(2015)]%
        {ryabova2015politeness}
\bibfield{author}{\bibinfo{person}{Marina Ryabova}.} \bibinfo{year}{2015}\natexlab{}.
\newblock \showarticletitle{Politeness strategy in everyday communication}.
\newblock \bibinfo{journal}{\emph{Procedia-Social and Behavioral Sciences}}  \bibinfo{volume}{206} (\bibinfo{year}{2015}), \bibinfo{pages}{90--95}.
\newblock


\bibitem[Samory and Mitra(2018a)]%
        {samory2018conspiracies}
\bibfield{author}{\bibinfo{person}{Mattia Samory} {and} \bibinfo{person}{Tanushree Mitra}.} \bibinfo{year}{2018}\natexlab{a}.
\newblock \showarticletitle{Conspiracies online: User discussions in a conspiracy community following dramatic events}. In \bibinfo{booktitle}{\emph{Proceedings of the International AAAI Conference on Web and Social Media}}, Vol.~\bibinfo{volume}{12}.
\newblock


\bibitem[Samory and Mitra(2018b)]%
        {samory2018government}
\bibfield{author}{\bibinfo{person}{Mattia Samory} {and} \bibinfo{person}{Tanushree Mitra}.} \bibinfo{year}{2018}\natexlab{b}.
\newblock \showarticletitle{'The Government Spies Using Our Webcams' The Language of Conspiracy Theories in Online Discussions}.
\newblock \bibinfo{journal}{\emph{Proceedings of the ACM on Human-Computer Interaction}} \bibinfo{volume}{2}, \bibinfo{number}{CSCW} (\bibinfo{year}{2018}), \bibinfo{pages}{1--24}.
\newblock


\bibitem[Siegel and Badaan(2020)]%
        {siegel2020no2sectarianism}
\bibfield{author}{\bibinfo{person}{Alexandra~A Siegel} {and} \bibinfo{person}{Vivienne Badaan}.} \bibinfo{year}{2020}\natexlab{}.
\newblock \showarticletitle{\# No2Sectarianism: Experimental approaches to reducing sectarian hate speech online}.
\newblock \bibinfo{journal}{\emph{American Political Science Review}} \bibinfo{volume}{114}, \bibinfo{number}{3} (\bibinfo{year}{2020}), \bibinfo{pages}{837--855}.
\newblock


\bibitem[Sunstein and Vermeule(2009)]%
        {sunstein2009conspiracy}
\bibfield{author}{\bibinfo{person}{Cass~R Sunstein} {and} \bibinfo{person}{Adrian Vermeule}.} \bibinfo{year}{2009}\natexlab{}.
\newblock \showarticletitle{Conspiracy theories: Causes and cures}.
\newblock \bibinfo{journal}{\emph{Journal of political philosophy}} \bibinfo{volume}{17}, \bibinfo{number}{2} (\bibinfo{year}{2009}), \bibinfo{pages}{202--227}.
\newblock


\bibitem[Tam and Chan(2023)]%
        {tam2023conspiracy}
\bibfield{author}{\bibinfo{person}{Kim-Pong Tam} {and} \bibinfo{person}{Hoi-Wing Chan}.} \bibinfo{year}{2023}\natexlab{}.
\newblock \showarticletitle{Conspiracy theories and climate change: A systematic review}.
\newblock \bibinfo{journal}{\emph{Journal of Environmental Psychology}} (\bibinfo{year}{2023}), \bibinfo{pages}{102129}.
\newblock


\bibitem[Treude and Storey(2009)]%
        {treude2009tagging}
\bibfield{author}{\bibinfo{person}{Christoph Treude} {and} \bibinfo{person}{Margaret-Anne Storey}.} \bibinfo{year}{2009}\natexlab{}.
\newblock \showarticletitle{How tagging helps bridge the gap between social and technical aspects in software development}. In \bibinfo{booktitle}{\emph{2009 IEEE 31st International Conference on Software Engineering}}. IEEE, \bibinfo{pages}{12--22}.
\newblock


\bibitem[Tuffley(2009)]%
        {tuffley2009mind}
\bibfield{author}{\bibinfo{person}{David Tuffley}.} \bibinfo{year}{2009}\natexlab{}.
\newblock \showarticletitle{Mind the gap}.
\newblock \bibinfo{journal}{\emph{International Journal of Sociotechnology and Knowledge Development (IJSKD)}} \bibinfo{volume}{1}, \bibinfo{number}{1} (\bibinfo{year}{2009}), \bibinfo{pages}{58--69}.
\newblock


\bibitem[Uscinski(2019)]%
        {uscinski2019conspiracy}
\bibfield{author}{\bibinfo{person}{Joseph~E Uscinski}.} \bibinfo{year}{2019}\natexlab{}.
\newblock \bibinfo{booktitle}{\emph{Conspiracy theories and the people who believe them}}.
\newblock \bibinfo{publisher}{Oxford University Press, USA}.
\newblock


\bibitem[Uscinski et~al\mbox{.}(2017)]%
        {uscinski2017climate}
\bibfield{author}{\bibinfo{person}{Joseph~E Uscinski}, \bibinfo{person}{Karen Douglas}, {and} \bibinfo{person}{Stephan Lewandowsky}.} \bibinfo{year}{2017}\natexlab{}.
\newblock \showarticletitle{Climate change conspiracy theories}.
\newblock In \bibinfo{booktitle}{\emph{Oxford research encyclopedia of climate science}}.
\newblock


\bibitem[Wofford(2022)]%
        {wofford2022parasitic}
\bibfield{author}{\bibinfo{person}{Morgan Wofford}.} \bibinfo{year}{2022}\natexlab{}.
\newblock \showarticletitle{Parasitic knowledge infrastructures: Data reuse by anthropogenic climate change skeptics}.
\newblock \bibinfo{journal}{\emph{Proceedings of the Association for Information Science and Technology}} \bibinfo{volume}{59}, \bibinfo{number}{1} (\bibinfo{year}{2022}), \bibinfo{pages}{837--839}.
\newblock


\bibitem[Wofford and Thomer(2023)]%
        {wofford2023curating}
\bibfield{author}{\bibinfo{person}{Morgan~F Wofford} {and} \bibinfo{person}{Andrea~K Thomer}.} \bibinfo{year}{2023}\natexlab{}.
\newblock \showarticletitle{Curating for Contrarian Communities: Data Practices of Anthropogenic Climate Change Skeptics}.
\newblock \bibinfo{journal}{\emph{Proceedings of the Association for Information Science and Technology}} \bibinfo{volume}{60}, \bibinfo{number}{1} (\bibinfo{year}{2023}), \bibinfo{pages}{442--455}.
\newblock


\end{thebibliography}
\newpage
\appendix
\section{Deploying Interventions on Reddit}
\label{appendix:lesson-learnt}
Deploying counternarratives in online communities, especially communities like r/climateskeptics that reject mainstream scientific consensus is challenging. Conspiracy theorists often distrust established institutions, including science, government, and media \cite{prooijen2018psychology}. They may view mainstream information sources as part of a conspiracy, making it challenging for researchers to introduce scientific evidence into the community. These challenges we encountered might provide important lessons for future research.

\subsection{Adapting interventions across subreddits}
As discussed in \autoref{section:data}, when deploying our intervention that originally designed for \textit{r/climateskeptics} in the community \textit{r/conspiracy}, some interventions misfired and were out of context (\autoref{fig:outofcontext-example}). This was because the two communities, though both discussed climate change conspiracy theories, used the same terms in different ways. For instance, as shown in \autoref{fig:outofcontext-example},  members of \textit{r/conspiracy} were discussing the use of ``chaos theory,'' but they did not refer to it in the context of climate change. This mismatch between the terminology and the context it was used in led to our intervention (about climate change) being misfired. Thus, future works exploring intervening in conspiracy theory-related communities should be aware of the uniqueness of each community, and adapt their interventions accordingly. This can help ensure the preciseness and effectiveness of any intervention deployed. 

\subsection{Account bans}
The Reddit accounts used for delivering interventions in the form of replies were transparently marked as bot accounts in the account description. Moreover, to assess the general community norms before starting the experiment, we ensured that both Reddit communities, r/climateskeptics, and r/conspiracy saw engagement from other bot accounts that are not banned. Despite this, several of our accounts got banned from r/climateskeptics without explanations. Our attempts to discuss the bans with community moderators were met with no response. Because of the bans, we could not deliver a large number of interventions to test the statistical significance of various intervention configurations on user responses.
We believe that the account bans may be a reflection of how the climate change denialist community may reject external attempts to discuss scientific evidence supporting climate change. We appealed several of our account bans and found that working with the moderation mechanism on Reddit is difficult, specifically in r/climateskeptics and r/conspiracy. It is possible that bots may not be deemed as trustworthy messengers of counter-attitudinal information, even when efforts were made to signal the bots' credibility with karma points.
Our experience highlights the importance of building trust between scientists and the online communities discussing conspiracy theories.

\subsection{Reddit API changes and Reddit Blackout}
In April 2023, Reddit announced changes in the API and data collection policies, which threatened the data access and access to moderation tools from third-party apps. Moderators of many subreddits leveraged their subreddits en masse to protest the API changes. Almost 7000 subreddits went private on June 12th notably including large subreddits such as r/funny, r/gaming, and r/science. The blackouts continued intermittently throughout June making for an unstable intervention environment. While r/climateskeptics didn't directly participate in the blackouts, the instability on the rest of the platform may have affected the traffic, participation, and general discussions on r/climateskeptics. We chose to continue the experiment throughout the blackouts because of uncertainty about the blackout timeline. While intervention experiments are best run in a stable social media environment, future researchers should consider intervention methods that are resilient to the constantly evolving landscape and monetization models of social media platforms.

\end{document}